\documentclass[aps,a4paper,showpacs,showkeys]{article}
\usepackage[twoside,top=2cm,bottom=2cm,left=2cm,right=2cm]{geometry}
\usepackage[utf8]{inputenc}	
\usepackage{authblk} 
\usepackage{epsfig}
\usepackage{amsmath}
\usepackage{amsfonts}
\usepackage{amssymb}
\usepackage{graphicx}
\usepackage{colordvi}
\usepackage[title]{appendix}
\usepackage{soul,color}

\date{}
\begin{document}
	\author{ Sergio De Filippo$^{(1)}$\footnote{%
			e-mail address: \textit{sdefilippo@sa.infn.it}}, Adele Naddeo$^{(2)}$\footnote{%
			e-mail address: \textit{anaddeo@na.infn.it}}}
	
\affil{\small{\textit{$^{(1)}$ Dipartimento di Fisica ``E. R. Caianiello", Universit\`{a} di Salerno, via Giovanni Paolo II, $84084$ Fisciano (SA), Italy}}}
\affil{\small{\textit{$^{(2)}$INFN, Sezione di Napoli, C. U. Monte S. Angelo, Via Cinthia, $80126$ Napoli, Italy}}}

\title{Microscopic foundation of thermodynamics, transition to classicality and regularization of gravitational-collapse singularities within Non-unitary $4$-th Derivative Gravity classically equivalent to Einstein gravity and its Newtonian limit}

\maketitle
	

\begin{abstract}
A detailed and updated review is given of De Filippo's Non-unitary $4$-th Derivative Gravity and its Newtonian limit, by pointing out the crucial role of non-unitarity in addressing transition to classicality and specifically localization of macroscopic bodies,
microscopic foundation of the second law of thermodynamics, measurement problem;
furthermore it provides a quantum field theory of gravity possibly not only
renormalizable but even finite, with a cancelation mechanism analogous to
supersymmetric field theories where cancelations are due to superpartners whereas
here to negative energy fields. Finally this non-unitary proposal addresses the
longstanding black hole information loss problem and this according to an unorthodox
view at variance with the mainstream endeavors to save unitarity at the expense of
changing General Relativity in vague unspecified ways. Last but not least motivations and conceptual framework are given, as the author
could not present them in his first papers written in a hurry since he was aware
that in a little time he would be unable to use pc keyboard or to write on paper due
to the progressing of motor neuron disease.
\end{abstract}

\tableofcontents
\newpage

\section{Introduction}

The discovery by S. Hawking that black holes radiate with a thermal spectrum corresponding to black hole temperature proportional to
horizon gravity leads to the conclusion that thermal radiation violates unitarity of quantum dynamics \cite{hawking1,hawking2}. In fact the result of the whole evaporation would be a highly mixed state of the radiation field, eventually pointing to a modification of ordinary quantum mechanics. This is the content of the so called Information Loss \footnote{The simplest way to understand information loss is considering that, when a very
massive star exhausts nuclear fusion and collapses into a black hole, all detailed
information on its composition and internal state are lost as the black hole is only
characterized by its mass, angular momentum and electric charge - plus other possible
conserved charges in models beyond the standard model - (black holes have no hair).} Paradox (for a review, see for instance \cite{mathur,unruh1,PolchinskiTASI} and references therein), which has been widely investigated throughout the last 40 years, triggering many endeavors in order to save unitarity.

In this respect a fundamental issue could be raised, i.e. why defend unitarity that is likely to be the source of the two weak points of the final setting of Quantum Mechanics (QM) by von Neumann \cite{vNeumann}:

\begin{itemize}
  \item measurement, based, as it is, on the vague notion of macroscopic measurement
apparatus, addressed but never solved by environment induced decoherence\cite{zurek,zurek1};
  \item coarse grained entropy, based on the poorly defined macroscopic observables\cite{coar}.
\end{itemize}

In fact both these two weak points can be addressed by a non-unitary quantum dynamics. This crucial observation lies at the basis of the unorthodox view expressed by one of us, SDF, in his model \cite{sergio1,sergio2,sergio3,sergio4,sergio5,sergio6}, which we explain in the present work with the aim of fully elucidating the underlying motivations and the resulting conceptual framework.

Indeed, the recent endeavors to save unitarity had as byproducts several interesting ideas like complementarity \cite{susskind1,susskind2,thooft1}, holographic principle \cite{bousso1,thooft2,susskind3}, AdS/CFT correspondence \cite{maldacena1,witten,sorkin,ryu,maldacena2,horowitz,hartman} and so on; but, for instance,
complementarity forced the introduction of a firewall changing the general relativity solution corresponding to a black hole \cite{fire1,fire2,fire3}, thus changing in a vague and unspecified way General Relativity (GR) itself, the most self-consistent and elegant physical
theory ever built since not based on experimental data like QM but on general principles. Furthermore anti de Sitter metric is cosmologically utterly irrelevant. One of the most interesting consequences of the quest for modifications of gravity in order to
save unitarity was the revival of Loop Quantum Gravity \cite{rovelli} that, according to SDF, is very stimulating even without reference to
quantization, since already at classical level loops of parallel transport in their difference from identity are an interesting alternative characterization of
curvature. These considerations lead to the suspect that trying to save unitarity at the expense of changing GR is like clutching at straws. This is not a solitary reckless idea if Polchinski, one of the main participants in the endeavor to save unitarity, in the introduction of his 2015 TASI lectures \cite{PolchinskiTASI} writes: \textit{But there is still the open question, where was Hawking's original argument wrong?}

As a result, the proposal here presented is based on the following requirements:

\begin{itemize}
 \item a fundamental change of quantum dynamics should not involve phenomenological
parameters at variance with collapse models \cite{ghirardi,pearle}, since $G$ (Newton constant),
$h$ (Planck constant) and $c$ (the speed of light) give a natural unit system by $l_p=\sqrt{Gh/c^3}$,  $m_p=\sqrt{hc/G}$,  $t_p=\sqrt{Gh/c^5}$,
and then any physical parameter is a pure number times a product of powers of $l_p$, $m_p$
and $t_p$, and this pure number is expected to be of the order of $10^0$;
 \item such a quantum dynamics should be non-unitary in order to address the microscopic
foundation of entropy and transition to classicality in closed systems, not to
mention black hole information problem and identification of the degrees of freedom
responsible for black hole entropy;
 \item in order to ensure consistency of such a non-unitary dynamics, it should come from
the unitary dynamics of a meta-system, composed of the physical system and an ancilla
\cite{kay,kay1}, so that once the ancilla is traced out a non-unitary evolution of the
physical system ensues, just like for open systems once the environment is traced out;
 \item this dynamics should come from a constrained theory in order to make non
observable ancilla degrees of freedom, as in every constrained theory (like Gupta
Bleuler's) the observable algebra is a subalgebra of the one defining the dynamics;
 \item this dynamics should be non Markov, at variance with models based on Lindblad
equation \cite{ghirardi1}, as it was argued long ago that a Markov non-unitary
dynamics would violate energy conservation \cite{unruh2}, whereas the present model keeps
energy expectation constant while introducing fluctuations leading to a quantum
microcanonical ensemble with non vanishing entropy, even starting from a pure state
with vanishing entropy \cite{entropy1,entropy2}.
\end{itemize}

The above requirements give rise to a model, whose non-unitary dynamics results from a unitary one of a meta-system, whose ancilla is a replica of the
physical system with the two subsystems of the meta-system interacting with each
other only through gravity. Such a dynamics includes both the traditional aspects of classical gravitational interactions and a kind of fundamental decoherence, which may be relevant to
the emergence of classical behavior of the center of mass of macroscopic bodies. The model in
fact treats on an equal footing mutual and self-interactions, which, according to some
authors, are believed to produce wave function localization and/or
reduction \cite{karolyhazy,frenkel1,frenkel,diosi0,diosi,penrose}. In particular, for ordinary condensed matter densities, it exhibits a localization
threshold at about $10^{11}$ proton masses, above which self-localized center of mass wave functions exist. Moreover an initial localized pure state evolves into a delocalized ensemble of localized states, characterized by an entropy slowly growing in time\cite{sergio2}. The latter circumstance is consistent with the expectation that, even for an isolated system, an entropy growth may eventually take place only due to the entanglement between observable and unobservable degrees of freedom via gravitational interaction. This feature has been tested, within the present framework, by taking an isolated homogeneous spherical macroscopic body of radius $R$ and mass $M$ above threshold \cite{sergio2,sergioFil1,sergioFil2}. Further preliminary results on a three-dimensional harmonic nanocrystal \cite{entropy2} show the ability of De Filippo's model to reproduce a gravity-induced relaxation towards thermodynamic equilibrium even for a closed system, in this way contributing to the long standing debate on the microscopic foundations of thermodynamics.

While the first formulation of De Filippo's model implies instantaneous action at a distance gravitational interactions between observable and unobservable degrees of freedom, in Ref.\cite{sergio3} it has been shown that a Hubbard-Stratonovich transformation of the gravitational interaction leads to a consistent field-theoretic interpretation, where the instantaneous interaction is replaced by a retarded potential. The result is the emergence of both a positive and a negative energy scalar field, each one coupled to the matter via a Yukawa interaction. The presence of a negative energy field has a natural interpretation within a dynamical theory implying wave function localization, in that it could supply the small continuous energy injection expected in such a kind of theories \cite{squires,ring}. As another interesting perspective, the emergence of negative energy fields naturally leads to the construction of finite field theories, where divergence cancelations take place thanks to the presence of couples of positive and negative energy fields, rather than of supersymmetric partners \cite{sergio3}. A further bonus of Hubbard-Stratonovich transformation is the proof that Newton-Schroedinger model \cite{penrose} can be obtained as a proper mean-field approximation\footnote{This reminds SDF of the attempts by G. 't Hooft to explore non perturbative aspects of QCD by promoting $SU(3)$ to $SU(N)$ sending $N \rightarrow \infty$, hoping to get a solvable
theory - that was afterwards proved to be perturbatively equivalent to (bosonic) string theory in $26$ dimensions - and then get information on finite $N$ values by $1/N$ expansion \cite{lucini}.} of De Filippo's model, although such an approximation does not share with the original model the key property of making linear superposition of macroscopically different states non observable \cite{sergio3,sergio4}. Last but not least, the presence of ghost fields suggests to look for a general covariant formulation within higher derivative gravity (HD), which has long been popular as a natural generalization of Einstein gravity \cite{dewitt,stelle,nojiri} although, already at the classical level, it is well known to be unstable due to the presence of runaway solutions \cite{hawkinghertog}.

These considerations lead to a non-unitary realization of HD gravity \cite{sergio5,sergio6}, which is classically stable and equivalent to Einstein gravity. It generalizes the previous non-relativistic formulation and shows a non-unitary Newtonian
limit compatible with the wavelike properties of microscopic particles and the classical behavior of center of mass of macroscopic bodies, as well as with a
trans-Planckian regularization of collapse singularities. That makes possible a unified reading of ordinary and black hole entropy as entanglement entropy with
hidden degrees of freedom \cite{unruh2,wald}.

One of us, SDF, wants to pay tribute to F. Karolyhazy who was the first, to the best
of his knowledge, to give quantitative estimates of gravitational decoherence \cite{karolyhazy,frenkel1,frenkel}.
His work, though largely unnoticed, was the first to investigate how GR could modify
QM, which minority SDF belongs to, whereas mainstream physicists were looking for
quantum gravity, i. e. how quantization modifies GR.

The plan of the present work is as follows.

In Section 2 an argument is presented, showing that for every system the corresponding ancilla has to be its replica with a symmetry constraint on the state space, in order for the ancilla not to be neither a source nor a sink of energy for the physical system \cite{unruh2}. Then it is shown that a characteristic feature of gravitational decoherence is the quasi-independence from mass density of the mass threshold for localization.

In Section 3 it is shown that a nonrelativistic model gives an explicit dynamics
realizing the expectations of Section 2, based on purely dimensional grounds; it is
also shown that the spreading of the probability density of the center of mass of a
macroscopic body is an entropic process. Furthermore  by means of a Hubbard-Stratonovich transformation it is shown that a
natural field theoretical reading includes negative energy fields suggesting a general covariant formulation in terms of HD gravity that is known to involve ghost fields. Finally it is shown that Newton-Schrodinger equation is just a mean field approximation of the present non-unitary non-relativistic model.

In Section 4 numerical and analytical tests of localization are reported.

In Section 5 numerical simulations are presented of two particles in a harmonic trap interacting with each other through an electrostatic delta like and ordinary gravitational interactions, and of an ideal harmonic nanocrystal; the last one shows a clear thermalization process towards a quantum microcanonical ensemble with its corresponding entropy.

In Section 6 it is shown that a non-unitary formulation of 4-th derivative gravity not only has as its newtonian limit the non-relativistic model of Section 3, but also gives in itself a Quantum Gravity possibly not only renormalizable but even finite. It is finally shown how non-unitary 4-th derivative gravity allows to regularize black hole singularities and to identify black hole entropy with von Neumann entropy of the matter inside the regularized singularity.

In Section 7 the issues of wave function reduction and alternative rigorous definition of coarse graining entropy are discussed.

Some concluding remarks and an outline of future perspectives end the paper.

\section{Localization in non relativistic quantum mechanics through non-unitary gravitational interaction}
\label{section2}

The two weak points of the final setting of Quantum Mechanics (QM) by von Neumann, the vague notion of a
macroscopic measurement apparatus and the definition of coarse graining entropy,
based, as it is, on the subjective notion of macroscopic observables, can in
principle both be addressed by a non-unitary quantum dynamics. The natural way to get
a non-unitary dynamics is to put the physical system in interaction with an ancillary
system and then tracing out the ancilla to get a mixed state described by a density
matrix, just as it happens with open systems when environment is traced out.
If we want to identify thermodynamic entropy with the entanglement one with an
ancillary or hidden system \cite{kay,kay1,unruh2}, we have to require that the hidden system neither to be a source nor a sink of
energy for the physical system and then at least to require thermal
equilibrium between the physical system and the hidden one:
\begin{equation}
\frac{1}{T} = \frac{\partial S}{\partial E_p}=\frac{\partial S}{\partial E_h}=\frac{\partial S}{\partial E_p}\left(\frac{\partial E_h}{\partial E_p} \right)^{-1},
\label{tequilibrium}
\end{equation}
where $E_p$ and $E_h$ denote respectively the energy of the physical and the hidden
system, while the entropy $S$ is one and the same for the two systems, being the
entanglement entropy of a bipartite system.

The above thermal equilibrium condition
leads to the almost inescapable conclusion that every physical system must have as
hidden partner its exact replica and the meta-system (physical plus hidden) state
space must be restricted by a symmetry constraint in the exchange of physical and
hidden degrees of freedom. This constraint eliminates the arbitrariness of
considering as non observable some degrees of freedom, since in any constrained
theory, like Gupta-Bleuler's, the observable algebra is a subalgebra of the original
dynamical one.

Several models were proposed to modify quantum dynamics in order to get localization
and transition to classicality, and some of them tried to establish a link with
gravity \cite{pearle1,rimini,ghirardi,diosi0,diosi}. They all introduced
phenomenological parameters, while for a fundamental theory $G$, $h$ and $c$ should be
enough since these three fundamental constants allow one to make all physical entities
dimensionless. If we limit ourselves to low energy physics, only $G$ and $h$ should appear in an
approximate low energy model and, for dimensional reasons, the only possible
relation between threshold mass $M$ and length $L$ for localization and transition to
classicality is
\begin{equation}
M^3 L = \frac{h^2}{G},
\label{thres1}
\end{equation}
which implies that
\begin{equation}
M = \left(\frac{h^2}{G} \right)^{\frac{3}{10}} \rho^{\frac{1}{10}}.
\label{thres2}
\end{equation}
Here one sees that the dependence on mass density $\rho$ is exceedingly weak: to get
a doubling of $M$, $\rho$ has to get $2^{10}=1024$ times higher. This quasi independence of
the threshold mass on mass density could be an experimental signature of gravitational
localization.

In Ref. \cite{sergio1} a dynamics was defined and analyzed which gives rise to
the same relation as above between threshold mass for localization and mass density. It
introduces a gravitational interaction only between a generic physical system and its replica and
constraints the meta-state space by a symmetry requirement as well. In this model no
gravitational interaction was introduced within physical and hidden system as the
author had in his mind the possibility to avoid gravitational-collapse
singularities.

To be specific, let $H[\psi ^{\dagger },\psi ]$ denote the second quantized
non-relativistic Hamiltonian of a finite number of particle species, like
electrons, nuclei, ions, atoms and/or molecules, according to the energy
scale. For notational simplicity $\psi ^{\dagger },\psi $ denote the whole
set $\psi _{j}^{\dagger }(x),\psi _{j}(x)$ of creation-annihilation
operators, i.e. one couple per particle species and spin component. This
Hamiltonian includes the usual electromagnetic interactions accounted for in
atomic and molecular physics. To incorporate gravitational interactions
including self-interactions, we introduce complementary
creation-annihilation operators $\chi _{j}^{\dagger }(x),\chi _{j}(x)$ and
the overall Hamiltonian
\begin{equation}
H_{G}=H[\psi ^{\dagger },\psi ]+H[\chi ^{\dagger },\chi
]-G\sum_{j,k}m_{j}m_{k}\int dxdy\frac{\psi _{j}^{\dagger }(x)\psi
_{j}(x)\chi _{k}^{\dagger }(y)\chi _{k}(y)}{|x-y|},
\end{equation}
acting on the tensor product $F_{\psi }\otimes F_{\chi }$ of the Fock spaces
of the $\psi $ and $\chi $ operators, where $m_{i}$ denotes the mass of the $%
i$-th particle species and $G$ is the gravitational constant. While the $%
\chi $ operators are taken to obey the same statistics as the original
operators $\psi $, we take advantage of the arbitrariness pertaining to
distinct fields and, for simplicity, we choose them commuting with one
another: $[\psi ,\chi ]$ $_{-}=[\psi ,\chi ^{\dagger }]_{-}=0$.

The meta-particle state space $S$ is identified with the subspace of $F_{\psi
}\otimes F_{\chi }$ including the meta-states obtained from the vacuum $%
\left| 0\right\rangle =\left| 0\right\rangle _{\psi }\otimes \left|
0\right\rangle _{\chi }$ by applying operators built in terms of the
products $\psi _{j}^{\dagger }(x)\chi _{j}^{\dagger }(y)$ and symmetrical
with respect to the interchange $\psi ^{\dagger }\leftrightarrow \chi
^{\dagger }$, which, as a consequence, have the same number of $\psi $
(green) and $\chi $ (red) meta-particles of each species. In particular for
instance the meta-states containing one green and one red $j$-meta-particle
are built by linear combinations of the symmetrized bilocal operators
\begin{equation}
\Phi _{j}^{\dagger }(x,y)\doteq \psi _{j}^{\dagger }(x)\chi _{j}^{\dagger
}(y)+\psi _{j}^{\dagger }(y)\chi _{j}^{\dagger }(x),
\end{equation}
by which the most general meta-state corresponding to one particle states is
represented by
\begin{equation}
\left| \left| f\right\rangle \right\rangle =\int dx\int dyf(x,y)\psi
_{j}^{\dagger }(x)\chi _{j}^{\dagger }(y)\left| 0\right\rangle
,\;\;f(x,y)=f(y,x).
\end{equation}
This is a consistent definition since the overall Hamiltonian is such that
the corresponding time evolution is a group of (unitary) endomorphisms of $S$. If we prepare a pure $n$-particle state, represented in the original
setting - excluding gravitational interactions - by
\begin{equation}
\left| g\right\rangle \doteq \int d^{n}xg(x_{1},x_{2},...,x_{n})\psi
_{j_{1}}^{\dagger }(x_{1})\psi _{j_{2}}^{\dagger }(x_{2})...\psi
_{j_{n}}^{\dagger }(x_{n})\left| 0\right\rangle ,
\end{equation}
its representation in $S$ is given by the meta-state
\begin{equation}
\left| \left| g\otimes g\right\rangle \right\rangle =\int
d^{n}xd^{n}yg(x_{1},...,x_{n})g(y_{1},...,y_{n})\psi _{j_{1}}^{\dagger
}(x_{1})...\psi _{j_{n}}^{\dagger }(x_{n})\chi _{j_{1}}^{\dagger
}(y_{1})...\chi _{j_{n}}^{\dagger }(y_{n})\left| 0\right\rangle .
\label{initial}
\end{equation}
As for the physical algebra, it is identified with the operator algebra of
say the green meta-world. In view of this, expectation values can be
evaluated by preliminarily tracing out the $\chi $ operators and then taking
the average in accordance with the traditional setting.

While we are talking trivialities as to an initial meta-state like in Eq. (\ref{initial}), that is not the case in the course of time, since the
overall Hamiltonian produces entanglement between the two meta-worlds,
leading, once $\chi $ operators are traced out, to mixed states of the
physical algebra. The ensuing non-unitary evolution induces both an
effective interaction mimicking gravitation, and wave function localization.

In fact, if we evaluate the time derivative of the linear
momentum, for notational simplicity for particles of one and the same type,
we get in the Heisenberg picture
\begin{eqnarray}
\frac{d\overrightarrow{p}}{dt} &=&-i\hslash \frac{d}{dt}\int dx\psi
^{\dagger }(x)\nabla \psi (x)\equiv \vec{F}+\vec{F}_{G}=  \nonumber \\
&&-\frac{i}{\hslash }\left[ \overrightarrow{p},H[\psi ^{\dagger },\psi ]%
\right] +Gm^{2}\int dx\psi ^{\dagger }(x)\psi (x)\nabla _{x}\int dy\frac{%
\chi ^{\dagger }(y)\chi (y)}{\left| x-y\right| }.  \label{dpdt}
\end{eqnarray}
If $\overrightarrow{p}$ denotes the total linear momentum, i.e. the $x$
integration extends to the whole space, and the expectation value in an
arbitrary meta-state vector of $S$ is considered, the gravitational force
vanishes, as it should be for self-gravitating matter, due to the
antisymmetry of the kernel $\nabla _{x}(1/\left| x-y\right| )$ and the
symmetry of the meta-states in the exchange $\psi ^{\dagger }\leftrightarrow
\chi ^{\dagger }$. On the other hand, if we evaluate the time derivative of
the linear momentum of a body contained in the space region $\Omega $, the
expectation of the corresponding gravitational force is
\begin{eqnarray}
\left\langle \left( \vec{F}_{G}\right) _{\Omega }\right\rangle  &\simeq
&Gm^{2}\left\langle \int\limits_{\Omega }dx\psi ^{\dagger }(x)\psi (x)\nabla
_{x}\int\limits_{\Omega }dy\frac{\chi ^{\dagger }(y)\chi (y)}{\left|
x-y\right| }\right\rangle   \nonumber \\
&&+Gm^{2}\int\limits_{\Omega }dx\left\langle \psi ^{\dagger }(x)\psi
(x)\right\rangle \nabla _{x}\int\limits_{R^{3}\backslash \Omega }dy\frac{%
\left\langle \chi ^{\dagger }(y)\chi (y)\right\rangle }{\left| x-y\right| }
\nonumber \\
&=&Gm^{2}\int\limits_{\Omega }dx\left\langle \psi ^{\dagger }(x)\psi
(x)\right\rangle \nabla _{x}\int\limits_{R^{3}\backslash \Omega }dy\frac{%
\left\langle \psi ^{\dagger }(x)\psi (x)\right\rangle }{\left| x-y\right| }.
\end{eqnarray}
The term referring to body self-interaction above vanishes once again due to
symmetry reasons, while in the following term long range correlations where
considered irrelevant as usual, and in the final result meta-state symmetry
was used once more. As for the center of mass coordinate, of course the
expression of its time derivative does not depend on gravitational
interactions at all.

This shows that the present model reproduces the classical aspects of the
naive theory without red meta-particles and with direct Coulomb-like
interactions between distinct particles only. On the other hand they
disagree as for the time dependence of phase coherences.

Consider in fact in the traditional gravitationless setting a physical body
in a given quantum state whose wave function $\Psi _{CM}(X)\Psi
_{INT}(x_{i}-x_{j})$ is the product of the wave function of the center of
mass and an internal stationary wave function dependent on a subset, for
instance, of the electronic and nuclear coordinates. In particular $\Psi
_{CM}$ can be chosen, for simplicity, in such a way that the corresponding
wave function in our model is itself, at least approximately, the product of
four factors: ${\it i}$)a wave function of the center of meta-mass, namely $%
(X+Y)/2$, where $Y$ is the center of mass of the corresponding red meta-body,
${\it ii}$)a stationary function of $X-Y$ describing the relative motion and
${\it iii}$) $\Psi _{INT}(x_{i}-x_{j})$ and ${\it iv}$)its red partner $\Psi
_{INT}(y_{i}-y_{j})$, namely:
\begin{eqnarray}
&&\Psi _{CM}(X)\Psi _{INT}(x_{i}-x_{j})\Psi _{CM}(Y)\Psi _{INT}(y_{i}-y_{j})
\nonumber \\
&=&\tilde{\Psi}_{CM}(\frac{X+Y}{2})\tilde{\Psi}_{INT}(X-Y)\Psi
_{INT}(x_{i}-x_{j})\Psi _{INT}(y_{i}-y_{j}).  \label{localized}
\end{eqnarray}
In Eq. (\ref{localized}) $\Psi _{INT}(x_{i}-x_{j})$ and its red partner are
obviously still stationary to an excellent approximation for not too large a
body mass $M$, since they are determined essentially from electromagnetic
interactions only. As to $\tilde{\Psi}_{INT}(X-Y)$ we choose it as the
ground state of the relative motion of the two interpenetrating meta-bodies,
which is formally equivalent to the plasma oscillations of two opposite
charge distributions. The corresponding potential energy, if the body is
spherically symmetric and not too far from being a homogeneous distribution
of radius $R,$ can be approximated for small relative displacements, on
purely dimensional grounds, by
\begin{equation}
U(X-Y)=\frac{1}{2}\alpha \frac{GM^{2}}{R^{3}}\left| X-Y\right| ^{2},
\end{equation}
where $\alpha \sim $ $10^{0}$ is a dimensionless constant. That means that
the relative ground state is represented by
\begin{equation}
\tilde{\Psi}_{INT}(X-Y)=\left( 2\Lambda ^{2}\pi \right) ^{-3/2}\exp \frac{%
-\left| X-Y\right| ^{2}}{2\Lambda ^{2}},\;\;\Lambda ^{2}=\frac{\hslash }{%
\sqrt{\alpha GM^{3}/R^{3}}}.  \label{gaussian}
\end{equation}
Then, if we choose $\Psi _{CM}(X)\propto \exp \left[ -(X/\Lambda )^{2}\right]
$, we get
\begin{equation}
\Psi _{CM}(X)\Psi _{CM}(Y)=\tilde{\Psi}_{INT}(X-Y)\tilde{\Psi}_{INT}(X+Y),
\end{equation}
with $\tilde{\Psi}_{INT}$ as in Eq. (\ref{gaussian}). In particular for
bodies of ordinary density $\sim 10^{24}m_{p}/cm^{3}$, where $m_{p}$ denotes
the proton mass, one gets
\begin{equation}
\Lambda \sim \sqrt{\frac{m_{p}}{M}} \ cm,
\end{equation}
which shows that the small displacement approximation is acceptable already
for $M\sim 10^{12}m_{p}$, when $\Lambda \sim 10^{-6}cm$, whereas the body
dimensions are $\sim 10^{-4}cm$.

If - in the traditional setting - we now consider at time $t=0$, omitting
the irrelevant factor $\Psi _{INT}$, a superposition
\begin{equation}
\frac{1}{\sqrt{2}}\left[ \Psi _{CM}(X)+\Psi _{CM}(X+Z)\right] ,
\end{equation}
the corresponding density matrix, before tracing out the red particle, is
represented in our model by
\begin{eqnarray}
&&\frac{1}{4}\left[ \bar{\Psi}_{CM}(X^{\prime })+\bar{\Psi}_{CM}(X^{\prime
}+Z)\right] \left[ \bar{\Psi}_{CM}(Y^{\prime })+\bar{\Psi}_{CM}(Y^{\prime
}+Z)\right]  \nonumber \\
&&\left[ \Psi _{CM}(X)+\Psi _{CM}(X+Z)\right] \left[ \Psi _{CM}(Y)+\Psi
_{CM}(Y+Z)\right] .
\end{eqnarray}
For not too long times and $Z\gtrsim 2R$, the main effect of time evolution
is due to the energy difference
\begin{equation}
E_{BIND}\simeq -GM^{2}/R\simeq 10^{-47}\left( \frac{M}{m_{p}}\right)
^{5/3}erg
\end{equation}
between products $\Psi _{CM}(X)\Psi _{CM}(Y)$, $\Psi _{CM}(X+Z)\Psi
_{CM}(Y+Z)$ corresponding to interpenetrating meta-bodies and $\Psi
_{CM}(X)\Psi _{CM}(Y+Z)$, $\Psi _{CM}(X+Z)\Psi _{CM}(Y)$, where the
gravitational interaction is irrelevant. After tracing out the red
meta-particles, we get at time $t$ the density matrix
\begin{eqnarray}
&&\frac{1}{2}\bar{\Psi}_{CM}(X^{\prime })\Psi _{CM}(X)+\frac{1}{2}\bar{\Psi}%
_{CM}(X^{\prime }+Z)\Psi _{CM}(X+Z)  \nonumber \\
&&+\frac{1}{2}\left[ \bar{\Psi}_{CM}(X^{\prime })\Psi _{CM}(X+Z)+\bar{\Psi}%
_{CM}(X^{\prime }+Z)\Psi _{CM}(X)\right] \cos \frac{E_{BIND}t}{\hslash },
\end{eqnarray}
leading to the emergence of classical behavior as soon as coherences for a
macroscopic body are unobservable due to their time oscillation. If for
instance we consider a body of $10^{21}$ proton masses, we get a frequency $%
E_{BIND}/\hslash \sim 10^{15}\sec ^{-1}$, corresponding to a length $\hslash
c/E_{BIND}\sim 10^{-5}cm$, much less than the radius $R\sim 10^{-1}cm$. This
shows that in such a case these coherences are totally unobservable as, in
order to detect them, a measurement time far lower than the time needed for
the light to cross the body would be needed.

Of course when the particle mass is not large enough for the oscillations to
hide coherences, they may decrease in time due to the difference in
spreading between the cases of interpenetrating and separated meta-bodies. In
the former case the spreading affects only the wave function of $X+Y$, while
in the latter the two gaussian wave packets for the two meta-bodies have
independent spreading. Since in general a gaussian wave packet $\propto \exp
[-x^{2}/(2\lambda ^{2})]$ spreads in time into a wave packet $\propto \exp
[-x^{2}/(2\lambda ^{2}+2i\hslash t/m)]$, where $m$ is the particle mass, the
typical time $\tau $ for the spreading to affect coherences is quite long:
\begin{equation}
\tau \sim \frac{M\Lambda ^{2}}{2\pi \hslash }\sim 10^{2}\sec .
\end{equation}

It is worthwhile remarking that, while the expression of the localization
length $\Lambda $ in Eq. (\ref{gaussian}) holds only for bodies whose mass is
not lower than say $10^{12}$ proton masses, another simple case corresponds
to masses lower than $10^{10}$ proton masses, where the relative motion
between the two meta-bodies in their ground state has not the character of
plasma oscillations any more, and they can be considered approximately as
point meta-particles. Their ground state wave function then becomes the
hydrogen-like wave function
\begin{equation}
\Psi (X-Y)\propto e^{-\left| X-Y\right| /a},\;\;a=\frac{2\hslash ^{2}}{GM^{3}%
}\sim 10^{25}\left( \frac{M}{m_{p}}\right) ^{-3}cm,
\end{equation}
which shows that the gravitational self-interaction between the green
meta-body and its red partner can be ignored for all practical purposes at
the molecular level.

As a result of the previous analysis, according to the present model,
fundamental decoherence due to gravitation is not expected to hide the
wavelike properties of particles even much larger than fullerene \cite{fullerene,fullerene1}, while it could still play a crucial role with
reference to the measurement problem in QM. While in fact environment
induced decoherence \cite{zurek,zurek1} can make it very hard to detect the usually
much weaker effects of fundamental decoherence, it cannot go farther than
produce entanglement with the environment. If and why such entangled states
should collapse is outside its scope.

\section{Non relativistic field theoretic setting and applications}
\label{section3}

In Section \ref{section2} a model for the gravitational interaction
in non relativistic quantum mechanics has been introduced. In it matter degrees of
freedom were duplicated and gravitational interactions were introduced
between observable and unobservable degrees of freedom only.

Interactions between observable and unobservable degrees of freedom are
instantaneous action at a distance ones and at first sight it looks unlikely
that they can be obtained as a non relativistic limit of more familiar local
interactions mediated by quantized fields. If, in fact, a local interaction
were introduced between an ordinary field and observable and unobservable
matter, the corresponding low energy limit would include an action at a
distance inside the observable (and the unobservable) matter too. In this
Section we want to show that a field theoretic reading of the model emerges
naturally through a Stratonovich-Hubbard transformation\cite{negele} of the
gravitational interaction. Within the minimal possible generalization of the
model, where the instantaneous interaction is replaced by a retarded
potential, the result of the transformation corresponds to the emergence of
an ordinary scalar field and a negative energy one, both coupled with the
matter through a Yukawa interaction. This result has a plain physical
reading, as it gives rise to two competing interactions with overall
vanishing effect within observable (and unobservable) matter: an attractive
and a repulsive one respectively mediated by the positive and the negative
energy field.

The presence of a negative energy field somehow is not surprising if we
consider that, in a dynamical theory supposed to account for wave function
localization, one expects a small continuous energy injection\cite
{squires,ring}, and negative energy fields are the most natural candidates
for that, apart from the possible introduction of a cosmological background.
In fact the possible role of negative-energy fields was already suggested
within some attempts to account for wave function collapse by
phenomenological stochastic models\cite{pearle}.

As a by-product of the Stratonovich-Hubbard transformation we show also that
a proper mean field approximation leads to the Schroedinger-Newton (S-N)
model \cite{kumar}. Finally, while the original model and the S-N
approximation are equivalent as to the classical aspects of the
gravitational interaction, the S-N model is shown to be ineffective in
turning off quantum coherences corresponding to different locations of one
and the same macroscopic body, at variance with the original model.

Let us adopt here an interaction representation, where the free Hamiltonian
is identified with $H[\psi ^{\dagger },\psi ]+H[\chi ^{\dagger },\chi ]$ and
the time evolution of an initially untangled meta-state $\left| \left| \tilde{%
\Phi}(0)\right\rangle \right\rangle $\ is represented by
\begin{eqnarray}
\left| \left| \tilde{\Phi}(t)\right\rangle \right\rangle &=&{\it T}\exp %
\left[ \frac{i}{\hslash }Gm^{2}\int dt\int dxdy\frac{\psi ^{\dagger
}(x,t)\psi (x,t)\chi ^{\dagger }(y,t)\chi (y,t)}{|x-y|}\right] \left| \left|
\tilde{\Phi}(0)\right\rangle \right\rangle  \nonumber \\
&\equiv &U(t)\left| \left| \tilde{\Phi}(0)\right\rangle \right\rangle \equiv
U(t)\left| \Phi (0)\right\rangle _{\psi }\otimes \left| \Phi
(0)\right\rangle _{\chi }.  \label{evolvedmetastate}
\end{eqnarray}
Then, by making use of a Stratonovich-Hubbard transformation\cite{negele},
we can rewrite the time evolution operator in the form
\begin{eqnarray}
U(t) &=&\int {\it D}\left[ \varphi _{1}\right] {\it D}\left[ \varphi _{2}%
\right] \exp \frac{ic^{2}}{2\hslash }\int dtdx\left[ \varphi _{1}\nabla
^{2}\varphi _{1}-\varphi _{2}\nabla ^{2}\varphi _{2}\right]  \nonumber \\
&&{\it T}\exp \left[ -i\frac{mc}{\hslash }\sqrt{2\pi G}\int dtdx\left[
\varphi _{1}(x,t)+\varphi _{2}(x,t)\right] \psi ^{\dagger }(x,t)\psi (x,t)%
\right]  \nonumber \\
&&{\it T}\exp \left[ -i\frac{mc}{\hslash }\sqrt{2\pi G}\int dtdx\left[
\varphi _{1}(x,t)-\varphi _{2}(x,t)\right] \chi ^{\dagger }(x,t)\chi (x,t)%
\right] ,  \label{stratonovich}
\end{eqnarray}
namely as a functional integral over two auxiliary real scalar fields $%
\varphi _{1}$ and $\varphi _{2}$.

To give a physical interpretation of this result, consider the minimal
variant of the Newton interaction in Eq. (\ref{evolvedmetastate}) aiming at
avoiding instantaneous action at a distance, namely consider replacing $%
-1/|x-y|$ by the Feynman propagator $4\pi \square ^{-1}\equiv 4\pi \left(
-\partial _{t}^{2}/c^{2}+\nabla ^{2}\right) ^{-1}$. Then the analog of Eq. (%
\ref{stratonovich}) holds with the d'Alembertian $\square $ replacing the
Laplacian $\nabla ^{2}$ and the ensuing expression can be read as the mixed
path integral and operator expression for the evolution operator
corresponding to the field Hamiltonian
\begin{eqnarray}
H_{Field} &=&H[\psi ^{\dagger },\psi ]+H[\chi ^{\dagger },\chi ]+\frac{1}{2}%
\int dx\left[ \pi _{1}^{2}+c^{2}\left| \nabla \varphi _{1}\right| ^{2}-\pi
_{2}^{2}-c^{2}\left| \nabla \varphi _{2}\right| ^{2}\right]
\label{fieldhamiltonian} \\
&&+mc\sqrt{2\pi G}\int dx\left\{ \left[ \varphi _{1}+\varphi _{2}\right]
\psi ^{\dagger }\psi +\left[ \varphi _{1}-\varphi _{2}\right] \chi ^{\dagger
}\chi \right\} ,
\end{eqnarray}
where $\pi _{1}=\dot{\varphi}_{1}$ and $\pi _{2}=\dot{\varphi}_{2}$
respectively denote the conjugate fields of $\varphi _{1}$ and $\varphi _{2}$
and all fields denote quantum operators. This theory can be read in analogy
with non relativistic quantum electrodynamics, where a relativistic field is
coupled with non relativistic matter, while the procedure to obtain the
corresponding action at a distance theory by integrating out the $\varphi $
fields is the analog of the Feynman's elimination of electromagnetic field
variables\cite{feynman}.

The resulting theory, containing the negative energy field $\varphi _{2}$,
has the attractive feature of being divergence free, at least in the
non-relativistic limit, where Feynman graphs with virtual
particle-antiparticle pairs can be omitted. To be specific, it does not
require the infinite self-energy subtraction needed for instance in
electrodynamics on evaluating the Lamb shift, or the coupling constant
renormalization\cite{itzykson}. Here of course we refer to the covariant
perturbative formalism applied to our model, where matter fields are
replaced by their relativistic counterparts and the non-relativistic
character of the model is reflected in the mass density being considered as
a scalar coupled with the scalar fields by Yukawa-like interactions. In fact
there is a complete cancellation among all Feynman diagrams containing only $%
\psi $ (or equivalently $\chi $) and internal $\varphi $ lines, owing to the
difference in sign between the $\varphi _{1}$ and the $\varphi _{2}$ free
propagators. This state of affairs of course is the field theoretic
counterpart of the absence of direct $\psi -\psi $ and $\chi -\chi $
interactions in the theory obtained by integrating out the $\varphi $
operators, whose presence would otherwise require the infinite self-energy
subtraction corresponding to normal ordering. These considerations,
supported by the mentioned suggestions derived from a phenomenological
analysis\cite{pearle} about the possible role of negative energy fields, may
provide substantial clues for the possible extensions of the model towards a
relativistic theory of gravity-induced localization. On the other hand a
Yukawa interaction with a (positive energy) scalar field emerges also moving
from Einstein's theory of gravitation, if one confines consideration to
conformal space-time fluctuations in a linear approximation \cite
{rosales,power}.

Going back to our evolved meta-state (\ref{evolvedmetastate}), the
corresponding physical state is given by
\begin{equation}
M(t)\equiv Tr_{\chi }\left| \left| \tilde{\Phi}(t)\right\rangle
\right\rangle \left\langle \left\langle \tilde{\Phi}(t)\right| \right|
=\sum_{k}\;\;\;_{\chi }\left\langle k\right| \left| \left| \tilde{\Phi}%
(t)\right\rangle \right\rangle \left\langle \left\langle \tilde{\Phi}%
(t)\right| \right| \left| k\right\rangle _{\chi },
\end{equation}
and, by using Eq. (\ref{stratonovich}), we can write
\begin{eqnarray}
_{\chi }\left\langle k\right| \left| \left| \tilde{\Phi}(t)\right\rangle
\right\rangle  &=&\int {\it D}\left[ \varphi _{1}\right] {\it D}\left[
\varphi _{2}\right] \exp \frac{ic^{2}}{2\hslash }\int dtdx\left[ \varphi
_{1}\nabla ^{2}\varphi _{1}-\varphi _{2}\nabla ^{2}\varphi _{2}\right]
\nonumber \\
&&_{\chi }\left\langle k\right| {\it T}\exp \left[ -i\frac{mc}{\hslash }%
\sqrt{2\pi G}\int dtdx\left[ \varphi _{1}(x,t)-\varphi _{2}(x,t)\right] \chi
^{\dagger }(x,t)\chi (x,t)\right] \left| \Phi (0)\right\rangle _{\chi }
\nonumber \\
&&{\it T}\exp \left[ -i\frac{mc}{\hslash }\sqrt{2\pi G}\int dtdx\left[
\varphi _{1}(x,t)+\varphi _{2}(x,t)\right] \psi ^{\dagger }(x,t)\psi (x,t)%
\right] \left| \Phi (0)\right\rangle _{\psi }.
\end{eqnarray}
Then the final expression for the physical state at time $t$ is given by
\begin{equation}
M(t)=  \label{exact}
\end{equation}
\begin{eqnarray*}
&&\int {\it D}\left[ \varphi _{1}\right] {\it D}\left[ \varphi _{2}\right]
{\it D}\left[ \varphi _{1}^{\prime }\right] {\it D}\left[ \varphi
_{2}^{\prime }\right] \exp \frac{ic^{2}}{2\hslash }\int dtdx\left[ \varphi
_{1}\nabla ^{2}\varphi _{1}-\varphi _{2}\nabla ^{2}\varphi _{2}-\varphi
_{1}^{\prime }\nabla ^{2}\varphi _{1}^{\prime }+\varphi _{2}^{\prime }\nabla
^{2}\varphi _{2}^{\prime }\right]  \\
&&_{\chi }\left\langle \Phi (0)\right| {\it T}^{-1}\exp \left[ i\frac{mc}{%
\hslash }\sqrt{2\pi G}\int dtdx\left[ \varphi _{1}^{\prime }-\varphi
_{2}^{\prime }\right] \chi ^{\dagger }\chi \right] {\it T}\exp \left[ -i%
\frac{mc}{\hslash }\sqrt{2\pi G}\int dtdx\left[ \varphi _{1}-\varphi _{2}%
\right] \chi ^{\dagger }\chi \right] \left| \Phi (0)\right\rangle _{\chi } \\
&&{\it T}\exp \left[ -i\frac{mc}{\hslash }\sqrt{2\pi G}\int dtdx\left[
\varphi _{1}+\varphi _{2}\right] \psi ^{\dagger }\psi \right] \left| \Phi
(0)\right\rangle _{\psi \psi }\left\langle \Phi (0)\right| {\it T}^{-1}\exp i%
\frac{mc}{\hslash }\sqrt{2\pi G}\int dtdx\left[ \varphi _{1}^{\prime
}+\varphi _{2}^{\prime }\right] \psi ^{\dagger }\psi ,
\end{eqnarray*}
where, due to the constraint on the meta-state space, $\chi $ operators can
be replaced by $\psi $ operators, if simultaneously the meta-state vector $%
\left| \Phi (0)\right\rangle _{\chi }$ is replaced by $\left| \Phi
(0)\right\rangle _{\psi }$. Then, if in the c-number factor corresponding to
the $\chi $-trace, we make the mean field (MF) approximation $\psi ^{\dagger
}\psi \rightarrow \left\langle \psi ^{\dagger }\psi \right\rangle $, we get
\[
M_{MF}(t)=
\]
\begin{eqnarray}
&&\int {\it D}\left[ \varphi _{1}\right] {\it D}\left[ \varphi _{2}\right]
{\it D}\left[ \varphi _{1}^{\prime }\right] {\it D}\left[ \varphi
_{2}^{\prime }\right] \exp \frac{ic^{2}}{2\hslash }\int dtdx\left[ \varphi
_{1}\nabla ^{2}\varphi _{1}-\varphi _{2}\nabla ^{2}\varphi _{2}-\varphi
_{1}^{\prime }\nabla ^{2}\varphi _{1}^{\prime }+\varphi _{2}^{\prime }\nabla
^{2}\varphi _{2}^{\prime }\right]   \nonumber \\
&&{\it T}\exp \left[ -i\frac{mc}{\hslash }\sqrt{2\pi G}\int dtdx\left[
\varphi _{1}\left[ \psi ^{\dagger }\psi +\left\langle \psi ^{\dagger }\psi
\right\rangle \right] +\varphi _{2}\left[ \psi ^{\dagger }\psi -\left\langle
\psi ^{\dagger }\psi \right\rangle \right] \right] \right] \left| \Phi
(0)\right\rangle _{\psi }  \nonumber \\
&&_{\psi }\left\langle \Phi (0)\right| {\it T}^{-1}\exp i\frac{mc}{\hslash }%
\sqrt{2\pi G}\int dtdx\left[ \varphi _{1}^{\prime }\left[ \psi ^{\dagger
}\psi +\left\langle \psi ^{\dagger }\psi \right\rangle \right] +\varphi
_{2}^{\prime }\left[ \psi ^{\dagger }\psi -\left\langle \psi ^{\dagger }\psi
\right\rangle \right] \right] .
\end{eqnarray}
Finally, if we perform functional integrations, we get
\begin{eqnarray}
M_{MF}(t) &=&  \nonumber \\
&&{\it T}\exp \left[ \frac{i}{\hslash }Gm^{2}\int dt\int dxdy\frac{\psi
^{\dagger }(x,t)\psi (x,t)\left\langle \psi ^{\dagger }(y,t)\psi
(y,t)\right\rangle }{|x-y|}\right] \left| \Phi (0)\right\rangle _{\psi }
\nonumber \\
&&_{\psi }\left\langle \Phi (0)\right| {\it T}^{-1}\exp \left[ -\frac{i}{%
\hslash }Gm^{2}\int dt\int dxdy\frac{\psi ^{\dagger }(x,t)\psi
(x,t)\left\langle \psi ^{\dagger }(y,t)\psi (y,t)\right\rangle }{|x-y|}%
\right] ,  \label{meanfield}
\end{eqnarray}
namely in this approximation the model is equivalent to the S-N theory\cite
{kumar}. Of course the whole procedure could be repeated without substantial
variations starting from the field theoretic Hamiltonian (\ref
{fieldhamiltonian}), inserting the mean field approximation before applying
the Feynman's procedure for the elimination of field variables, and then
getting a retarded potential version of the S-N model.

However this approximation does not share with the original model the
crucial ability of making linear superpositions of macroscopically different
states unobservable. Consider in fact an initial state corresponding to the
linear, for simplicity orthogonal, superposition of $N$ localized states of
one and the same macroscopic body, which were shown to exist as pure states
corresponding to unentangled bound meta-states of green and red meta-matter
for bodies of ordinary density and a mass higher than $\sim 10^{11}$ proton
masses\cite{sergio1}:
\begin{equation}
\left| \Phi (0)\right\rangle =\frac{1}{\sqrt{N}}\sum_{j=1}^{N}\left|
z_{j}\right\rangle ,  \label{superposition}
\end{equation}
where $\left| z\right\rangle $ represents a localized state centered in $z$.
Compare the coherence $\left\langle z_{h}\right| M(t)\left|
z_{k}\right\rangle $ when evaluated according to Eqs. (\ref{exact}) and (\ref
{meanfield}), where we consider the localized states as approximate
eigenstates of the particle density operator
\begin{equation}
\psi ^{\dagger }(x,t)\psi (x,t)\left| z\right\rangle \simeq n(x-z)\left|
z\right\rangle
\end{equation}
and time dependence in $\psi ^{\dagger }\psi $ irrelevant, as the considered
states are stationary states in the Schroedinger picture apart from an
extremely slow spreading\cite{sergio2}.

According to the original model we get, from Eq. (\ref{exact})
\begin{eqnarray}
&&\left\langle z_{h}\right| M(t)\left| z_{k}\right\rangle  \nonumber \\
&=&\int {\it D}\left[ \varphi _{1}\right] {\it D}\left[ \varphi _{2}\right]
{\it D}\left[ \varphi _{1}^{\prime }\right] {\it D}\left[ \varphi
_{2}^{\prime }\right] \exp \frac{ic^{2}}{2\hslash }\int dtdx\left[ \varphi
_{1}\nabla ^{2}\varphi _{1}-\varphi _{2}\nabla ^{2}\varphi _{2}-\varphi
_{1}^{\prime }\nabla ^{2}\varphi _{1}^{\prime }+\varphi _{2}^{\prime }\nabla
^{2}\varphi _{2}^{\prime }\right]  \nonumber \\
&&\frac{1}{N^{2}}\sum_{j=1}^{N}\exp \left[ -i\frac{mc}{\hslash }\sqrt{2\pi G}%
\int dtdx\left[ \left[ \varphi _{1}-\varphi _{2}\right] n(x-z_{j})-\left[
\varphi _{1}^{\prime }-\varphi _{2}^{\prime }\right] n(x-z_{j})\right] %
\right]  \nonumber \\
&&\exp \left[ -i\frac{mc}{\hslash }\sqrt{2\pi G}\int dtdx\left[ \left[
\varphi _{1}+\varphi _{2}\right] n(x-z_{h})-\left[ \varphi _{1}^{\prime
}+\varphi _{2}^{\prime }\right] n(x-z_{k})\right] \right] ,
\end{eqnarray}
and, after integrating out the scalar fields,
\begin{equation}
\left\langle z_{h}\right| M(t)\left| z_{k}\right\rangle =\frac{1}{N^{2}}%
\sum_{j=1}^{N}\exp \frac{i}{\hslash }Gm^{2}t\int dxdy\left[ \frac{%
n(x-z_{j})n(y-z_{h})}{|x-y|}-\frac{n(x-z_{j})n(y-z_{k})}{|x-y|}\right] ,
\label{coherences}
\end{equation}
which shows that, while diagonal coherences are given by $\left\langle
z_{h}\right| M(t)\left| z_{h}\right\rangle =1/N$, the off-diagonal ones,
under reasonable assumptions on the linear superposition in Eq. (\ref
{superposition}) of a large number of localized states, approximately
vanish, due to the random phases in the sum in Eq. (\ref{coherences}). This
makes the state $M(t)$, for not too short times, equivalent to an ensemble
of localized states:
\begin{equation}
M(t)\simeq \frac{1}{N}\sum_{j=1}^{N}\left| z_{j}\right\rangle \left\langle
z_{j}\right| .
\end{equation}
On the other hand, if we calculate coherences according to the S-N model, we
get
\begin{eqnarray}
&&\left\langle z_{h}\right| M_{MF}(t)\left| z_{k}\right\rangle  \nonumber \\
&=&\frac{1}{N}\exp \frac{i}{\hslash }Gm^{2}t\int dxdy\frac{\left[
n(x-z_{h})-n(x-z_{k})\right] \sum_{j=1}^{N}n(y-z_{j})/N}{|x-y|},
\end{eqnarray}
so that here the sum appears in the exponent and there is no cancellation.
While diagonal coherences keep the same value as in the original model,
off-diagonal ones acquire only a phase for the presence of the mean
gravitational interaction, but keep the same absolute value $1/N$ as the
diagonal ones. Furthermore, if we take $N=2$ rather than $N$ very large, the
S-N approximation gives just $\left\langle z_{1}\right| M_{MF}(t)\left|
z_{2}\right\rangle =1/2$, whereas the exact model still offers a mechanism
to make off-diagonal coherences unobservable, due to time oscillations\cite
{sergio1}.

It is worth while to remind that, although both our proposal and the S-N
model give rise to localized states and reproduce the classical aspects of
the gravitational interaction, the mean field approximation, necessary to
pass from the former to the latter, spoils the theory, not only of its
feature of reducing unlocalized wave functions, but also of another
desirable property. In fact it can be shown that according to our model
localized states evolve into unlocalized ensembles of localized states\cite
{sergio2} (see Section \ref{section2}), while the S-N theory leads to stationary localized states\cite
{kumar}, which is rather counterintuitive and unphysical, since space
localization implies linear momentum uncertainty, and this, in its turn,
should imply a spreading of the probability distribution in space.

In conclusion, while the present model has only a non-relativistic character,
its analysis hints of possible directions for extensions to higher energies,
where an instantaneous action at a distance is not appropriate. In
particular, the emergence of negative energy fields leads naturally to a
promising perspective for the construction of finite field theories, where
divergence cancellations are due to the presence of couples of positive and
negative energy fields, rather than of supersymmetric partners. Furthermore,
since the geometric formulation of Newtonian gravity, i.e. the Newton-Cartan
theory, leads only to the mean field approximation, i.e. the S-N theory\cite
{christian}, it may be likely that the geometric aspects of gravity may even
play a misleading role in looking for a quantum theory including gravity
both in its classical aspects and in its possible localization effects. More
specifically the Einstein theory of gravitation could arise, unlike, for
instance, classical electrodynamics, not as a result of taking expectation
values with respect to a pure physical state, but rather, as an effective
long distance theory like hydrodynamics, from a statistical average, or
equivalently by tracing out unobservable degrees of freedom starting from a
pure meta-state.

\subsection{Generalization to $N$ colors and $N \rightarrow \infty$ limit}

The aim of the present subsection is to give a well defined procedure for passing
from the original model to the SN approximation, replacing the heuristic
suggestion given above \cite{sergio4}. In doing that, the original model with
just two colors, green and red, is considered as the simplest
representative, for $N=2$, of a class of $N$-color models, whereas the SN
model is obtained as the $N\rightarrow \infty $ limit. While clarifying the
relationship between our proposal and the SN model, this result gives in
principle the possibility to develop $1/N$ expansions in analogy to what is
done in ordinary condensed matter physics\cite{stanley,largeN}. In
particular, while the $N\rightarrow \infty $ limit does not involve
decoherence, but only localization, $1/N$ expansions may provide approximate
schemes for the evaluation of decoherence.

To be specific, following Section \ref{section2}, let $H[\psi ^{\dagger
},\psi ]$ denote the second quantized non-relativistic Hamiltonian of a
finite number of particle species, like electrons, nuclei, ions, atoms
and/or molecules, according to the energy scale. For notational simplicity $%
\psi ^{\dagger },\psi $ denote the whole set $\psi _{j}^{\dagger }(x),\psi
_{j}(x)$ of creation-annihilation operators, i.e. one couple per particle
species and spin component. This Hamiltonian includes the usual
electromagnetic interactions accounted for in atomic and molecular physics.
To incorporate gravitational interactions including self-interactions, we
introduce a color quantum number $\alpha =1,2,...,N$, in such a way that
each couple $\psi _{j}^{\dagger }(x),\psi _{j}(x)$ is replaced by $N$
couples $\psi _{j,\alpha }^{\dagger }(x),\psi _{j,\alpha }(x)$ of
creation-annihilation operators. The overall Hamiltonian, including
gravitational interactions and acting on the tensor product $%
\bigotimes_{\alpha }F_{\alpha }$ of the Fock spaces of the $\psi _{\alpha }$
operators, is then given by
\begin{equation}
H_{G}=\sum_{\alpha =1}^{N}H[\psi _{\alpha }^{\dagger },\psi _{\alpha }]-
\frac{G}{N-1}\sum_{j,k}m_{j}m_{k}\sum_{\alpha <\beta }\int dxdy\frac{\psi
_{j,\alpha }^{\dagger }(x)\psi _{j,\alpha }(x)\psi _{k,\beta }^{\dagger
}(y)\psi _{k,\beta }(y)}{|x-y|},
\end{equation}
where here and henceforth Greek indices denote color indices, $\psi _{\alpha
}\equiv (\psi _{1,\alpha }\psi _{2,\alpha },...\psi _{N,\alpha })$, and $%
m_{i}$ denotes the mass of the $i$-th particle species, while $G$ is the
gravitational constant.

While the $\psi _{\alpha }$ operators obey the same statistics as the
original operators $\psi $, we take advantage of the arbitrariness
pertaining to distinct operators and, for simplicity, we choose them
commuting with one another: $\alpha \neq \beta \Rightarrow $ $[\psi _{\alpha
},\psi _{\beta }]$ $_{-}=[\psi _{\alpha },\psi _{\beta }^{\dagger }]_{-}=0$.
The meta-particle state space $S$ is identified with the subspace of $%
\bigotimes_{\alpha }F_{\alpha }$ including the meta-states obtained from the
vacuum $\left| \left| 0\right\rangle \right\rangle =\bigotimes_{\alpha
}\left| 0\right\rangle _{\alpha }$ by applying operators built in terms of
the products $\prod_{\alpha =1}^{N}\psi _{j,\alpha }^{\dagger }(x_{\alpha })$
and symmetrical with respect to arbitrary permutations of the color indices,
which, as a consequence, for each particle species, have the same number of
meta-particles of each color. This is a consistent definition since the time
evolution generated by the overall Hamiltonian is a group of (unitary)
endomorphisms of $S$. If we prepare a pure $n$-particle state, represented
in the original setting - excluding gravitational interactions - by
\begin{equation}
\left| g\right\rangle \doteq \int d^{n}xg(x_{1},x_{2},...,x_{n})\psi
_{j_{1}}^{\dagger }(x_{1})\psi _{j_{2}}^{\dagger }(x_{2})...\psi
_{j_{n}}^{\dagger }(x_{n})\left| 0\right\rangle ,
\end{equation}
its representative in $S$ is given by the meta-state
\begin{equation}
\left| \left| g^{\otimes N}\right\rangle \right\rangle \doteq \prod_{\alpha
} \left[ \int d^{n}xg(x_{1},x_{2},...,x_{n})\psi _{j_{1},\alpha }^{\dagger
}(x_{1})...\psi _{j_{n},\alpha }^{\dagger }(x_{n})\right] \left| \left|
0\right\rangle \right\rangle .  \label{initial1}
\end{equation}
As for the physical algebra, it is identified with the operator algebra of
say the $\alpha =1$ meta-world. In view of this, expectation values can be
evaluated by preliminarily tracing out the unobservable operators, namely
with $\alpha >1$, and then taking the average of an operator belonging to
the physical algebra. It should be made clear that we are not prescribing an
ad hoc restriction of the observable algebra. Once the constraint
restricting $\bigotimes_{\alpha }F_{\alpha }$ to $S$ is taken into account,
in order to get an effective gravitational interaction between particles of
one and the same color\cite{sergio1}, the resulting state space does not
contain states that can distinguish between operators of different color.
The only way to accommodate a faithful representation of the physical
algebra within the meta-state space is to restrict the algebra.

While we are talking trivialities as to an initial meta-state like in Eq. (%
\ref{initial1}), that is not the case in the course of time, since the
overall Hamiltonian produces entanglement between meta-worlds of different
color, leading, once unobservable operators are traced out, to mixed states
of the physical algebra. A peculiar feature of the model is that it cannot
be obtained by quantizing its naive classical version, since the classical
states corresponding to the constraint in $\bigotimes_{\alpha }F_{\alpha }$,
selecting the meta-state space $S$, have partners of all colors sitting in
one and the same space point and then a divergent gravitational energy.
While it is usual that, in passing from the classical to the quantum
description, self-energy divergences are mitigated, in this instance we pass
from a completely meaningless classical theory to a quite divergence free
one. This is more transparent in a field theoretic description\cite
{sergio3}.

Let us adopt here an interaction representation, where the free Hamiltonian
is identified with $\sum_{\alpha =1}^{N}H[\psi _{\alpha }^{\dagger },\psi
_{\alpha }]$ and the time evolution of an initially unentangled meta-state $%
\left| \left| \tilde{\Phi}(0)\right\rangle \right\rangle =\bigotimes_{\alpha
=1}^{N}\left| \Phi (0)\right\rangle _{\alpha }$\ is represented by
\begin{eqnarray}
\left| \left| \tilde{\Phi}(t)\right\rangle \right\rangle &=&{\it T}\exp %
\left[ \frac{iG}{(N-1)\hslash }m^{2}\sum_{\alpha <\beta }\int dt\int dxdy%
\frac{\psi _{\alpha }^{\dagger }(x,t)\psi _{\alpha }(x,t)\psi _{\beta
}^{\dagger }(y,t)\psi _{\beta }(y,t)}{|x-y|}\right] \left| \left| \tilde{\Phi%
}(0)\right\rangle \right\rangle  \nonumber \\
&\equiv &U(t)\left| \left| \tilde{\Phi}(0)\right\rangle \right\rangle ,
\label{evolvedmetastate1}
\end{eqnarray}
where for notational simplicity we are referring to just one particle
species.

Then, by using a Stratonovich-Hubbard transformation\cite{negele}, we can
rewrite $U(t)$ as
\begin{eqnarray}
U(t) &=&\int {\it D}\left[ \varphi \right] \prod_{\alpha }{\it D}\left[
\varphi _{\alpha }\right] \exp \frac{ic^{2}}{2\hslash }\int dtdx\left[
\varphi (x,t)\nabla ^{2}\varphi (x,t)-\sum_{\alpha }\varphi _{\alpha
}(x,t)\nabla ^{2}\varphi _{\alpha }(x,t)\right]  \nonumber \\
&&{\it T}\exp \left[ -i\frac{2mc}{\hslash }\sqrt{\frac{\pi G}{N-1}}
\sum_{\alpha }\int dtdx\left[ \varphi (x,t)+\varphi _{\alpha }(x,t)\right]
\psi _{\alpha }^{\dagger }(x,t)\psi _{\alpha }(x,t)\right] ,
\label{stratonovich1}
\end{eqnarray}
i.e. as a functional integral over the auxiliary real scalar fields $\varphi
,\varphi _{1},\varphi _{2},...,\varphi _{N}$.

Just as for $N=2$ case, a physical interpretation of this result
can be given, by considering the minimal variant of the Newton interaction
in Eq. (\ref{evolvedmetastate1}) aiming at avoiding instantaneous action at a
distance, namely replacing $-1/|x-y|$ by the Feynman propagator $4\pi
\square ^{-1}\equiv 4\pi \left( -\partial _{t}^{2}/c^{2}+\nabla ^{2}\right)
^{-1}$. Then the analog of Eq. (\ref{stratonovich1}) holds with the
d'Alembertian $\square $ replacing the Laplacian $\nabla ^{2}$ and the
ensuing expression can be read as the mixed path integral and operator
expression for the evolution operator corresponding to the field Hamiltonian
\begin{eqnarray}
H_{Field} &=&\sum_{\alpha =1}^{N}H[\psi _{\alpha }^{\dagger },\psi _{\alpha
}]+\frac{1}{2}\int dx\left[ \pi ^{2}+c^{2}\left| \nabla \varphi \right|
^{2}-\sum_{\alpha =1}^{N}\left( \pi _{\alpha }^{2}+c^{2}\left| \nabla
\varphi _{\alpha }\right| ^{2}\right) \right]  \nonumber \\
&&+2mc\sqrt{\frac{\pi G}{N-1}}\int dx\sum_{\alpha =1}^{N}\left\{ \left[
\varphi +\varphi _{\alpha }\right] \psi _{\alpha }^{\dagger }\psi _{\alpha
}\right\} ,
\end{eqnarray}
where $\pi =\dot{\varphi}$ and $\pi _{\alpha }=\dot{\varphi}_{\alpha }$
respectively denote the conjugate fields of $\varphi $ and $\varphi _{\alpha
}$ and all fields are quantum operators. This can be read in analogy with
non relativistic quantum electrodynamics, where a relativistic field is
coupled with non relativistic matter, while the procedure to obtain the
corresponding action at a distance theory by integrating out the $\varphi $
fields is the analog of the Feynman's elimination of electromagnetic field
variables\cite{feynman}.

The resulting theory, containing the negative energy fields $\varphi
_{\alpha }$, has the attractive feature of being divergence free, at least
in the non-relativistic limit, where Feynman graphs with virtual
particle-antiparticle pairs can be omitted. To be specific, it does not
require the infinite self-energy subtraction needed for instance in
electrodynamics on evaluating the Lamb shift, or the coupling constant
renormalization\cite{itzykson}. Here of course we refer to the covariant
perturbative formalism applied to our model, where matter fields are
replaced by their relativistic counterparts and the non-relativistic
character of the model is reflected in the mass density being considered as
a scalar coupled with the scalar fields by Yukawa-like interactions. In fact
there is a complete cancellation, for a fixed color index $\alpha $, among
all Feynman diagrams containing only $\psi _{\alpha }$ and internal $\varphi
$ and $\varphi _{\alpha }$ lines, owing to the difference in sign between
the $\varphi $ and the $\varphi _{\alpha }$ free propagators. This state of
affairs of course is the field theoretic counterpart of the absence of
direct $\psi _{\alpha }-\psi _{\alpha }$ interactions in the theory obtained
by integrating out the $\varphi $ operators, whose presence would otherwise
require the infinite self-energy subtraction corresponding to normal
ordering.

Going back to our meta-state (\ref{evolvedmetastate1}), the partial trace
\begin{equation}
M(t)\equiv \stackrel{\circ }{Tr}\left| \left| \tilde{\Phi}(t)\right\rangle
\right\rangle \left\langle \left\langle \tilde{\Phi}(t)\right| \right|
\equiv \sum_{k_{2},k_{3},...,k_{N}}\;\;\;\left\langle \left\langle
k_{2},k_{3},...,k_{N}\right| \right| \left| \left| \tilde{\Phi}
(t)\right\rangle \right\rangle \left\langle \left\langle \tilde{\Phi}
(t)\right| \right| \left| \left| k_{2},k_{3},...,k_{N}\right\rangle
\right\rangle ,
\end{equation}
gives the corresponding physical state, where $\left| \left|
k_{2},k_{3},...,k_{N}\right\rangle \right\rangle \equiv \bigotimes_{\alpha
=2}^{N}\left| k_{\alpha }\right\rangle _{\alpha }$ and $(\left|
k\right\rangle _{\alpha })_{k=1,2,...}$ denotes an orthonormal basis in the
Fock space $F_{\alpha }$. Then, by using Eq. (\ref{stratonovich1}), we can
write
\[
M(t)=
\]
\begin{eqnarray}
&&\int {\it D}\left[ \varphi \right] \prod_{\alpha }{\it D}\left[ \varphi
_{\alpha }\right] {\it D}\left[ \varphi ^{\prime }\right] \prod_{\alpha }
{\it D}\left[ \varphi _{\alpha }^{\prime }\right] \exp \frac{ic^{2}}{
2\hslash }\int dtdx\left[ \varphi \nabla ^{2}\varphi -\sum_{\alpha }\varphi
_{\alpha }\nabla ^{2}\varphi _{\alpha }-\varphi ^{\prime }\nabla ^{2}\varphi
^{\prime }+\sum_{\alpha }\varphi _{\alpha }^{\prime }\nabla ^{2}\varphi
_{\alpha }^{\prime }\right]  \nonumber \\
&&\left[ \bigotimes_{\alpha =2}^{N}\;_{\alpha }\left\langle \Phi (0)\right| %
\right] {\it T}^{-1}\exp \left[ i\frac{2mc}{\hslash }\sqrt{\frac{\pi G}{N-1}}
\sum_{\alpha =2}^{N}\int dtdx\left[ \varphi ^{\prime }(x,t)+\varphi _{\alpha
}^{\prime }(x,t)\right] \psi _{\alpha }^{\dagger }(x,t)\psi _{\alpha }(x,t) %
\right]  \nonumber \\
&&{\it T}\exp \left[ -i\frac{2mc}{\hslash }\sqrt{\frac{\pi G}{N-1}}
\sum_{\alpha =2}^{N}\int dtdx\left[ \varphi (x,t)+\varphi _{\alpha }(x,t) %
\right] \psi _{\alpha }^{\dagger }(x,t)\psi _{\alpha }(x,t)\right]
\bigotimes_{\alpha =2}^{N}\left| \Phi (0)\right\rangle _{\alpha }  \nonumber
\\
&&{\it T}\exp \left[ -i\frac{2mc}{\hslash }\sqrt{\frac{\pi G}{N-1}}\int dtdx %
\left[ \varphi (x,t)+\varphi _{1}(x,t)\right] \psi _{1}^{\dagger }(x,t)\psi
_{1}(x,t)\right] \left| \Phi (0)\right\rangle _{1}  \nonumber \\
&&_{1}\left\langle \Phi (0)\right| {\it T}^{-1}\exp \left[ i\frac{2mc}{
\hslash }\sqrt{\frac{\pi G}{N-1}}\int dtdx\left[ \varphi ^{\prime
}(x,t)+\varphi _{1}^{\prime }(x,t)\right] \psi _{1}^{\dagger }(x,t)\psi
_{1}(x,t)\right] .
\end{eqnarray}

To study the $N\rightarrow \infty $ limit of this expression, replace the
products $\varphi \psi _{1}^{\dagger }\psi _{1}$ and $\varphi ^{\prime }\psi
_{1}^{\dagger }\psi _{1}$ respectively with $\tilde{\varphi}\psi
_{1}^{\dagger }\psi _{1}$ and $\tilde{\varphi}^{\prime }\psi _{1}^{\dagger
}\psi _{1}$, inserting simultaneously the $\delta $-functionals $\delta
\lbrack \varphi -\tilde{\varphi}]=\int {\it D}\left[ g\right] \exp (i\frac{%
2mc}{\hslash }\sqrt{\frac{\pi G}{N-1}}\int dtdxg[\tilde{\varphi}-\varphi ])$%
, $\delta \lbrack \varphi ^{\prime }-\tilde{\varphi}^{\prime }]=\int {\it D}%
\left[ g^{\prime }\right] \exp (i\frac{2mc}{\hslash }\sqrt{\frac{\pi G}{N-1}}%
\int dtdxg^{\prime }[\varphi ^{\prime }-\tilde{\varphi}^{\prime }])$. Then
we can perform functional integration on $\varphi $,$\varphi ^{\prime }$ and
$\varphi _{\alpha }$,$\varphi _{\alpha }^{\prime }$ for $\alpha =2,3,...,N$
and get
\[
M(t)=
\]
\begin{eqnarray}
&&\int {\it D}\left[ g\right] {\it D}\left[ g^{\prime }\right] \left[
\bigotimes_{\alpha =2}^{N}\;_{\alpha }\left\langle \Phi (0)\right| \right]
\nonumber \\
&&{\it T}^{-1}\exp \frac{-iG}{\hslash }m^{2}\int dt\int dxdy\left[
\sum_{1<\alpha <\beta }\frac{\psi _{\alpha }^{\dagger }(x)\psi _{\alpha
}(x)\psi _{\beta }^{\dagger }(y)\psi _{\beta }(y)}{(N-1)|x-y|}+\frac{\left[
\sum_{1<\alpha }\psi _{\alpha }^{\dagger }(x)\psi _{\alpha }(x)+g^{\prime
}(x)/2\right] g^{\prime }(y)}{(N-1)|x-y|}\right]  \nonumber \\
&&{\it T}\exp \frac{iG}{\hslash }m^{2}\int dt\int dxdy\left[ \sum_{1<\alpha
<\beta }\frac{\psi _{\alpha }^{\dagger }(x)\psi _{\alpha }(x)\psi _{\beta
}^{\dagger }(y)\psi _{\beta }(y)}{(N-1)|x-y|}+\frac{\left[ \sum_{1<\alpha
}\psi _{\alpha }^{\dagger }(x)\psi _{\alpha }(x)+g(x)/2\right] g(y)}{%
(N-1)|x-y|}\right]  \nonumber \\
&&\bigotimes_{\alpha =2}^{N}\left| \Phi (0)\right\rangle _{\alpha }\int {\it %
D}\left[ \tilde{\varphi}\right] {\it D}\left[ \tilde{\varphi}^{\prime }%
\right] {\it D}\left[ \varphi _{1}\right] {\it D}\left[ \varphi _{1}^{\prime
}\right] \exp \frac{ic^{2}}{2\hslash }\int dtdx\left[ -\varphi _{1}\nabla
^{2}\varphi _{1}+\varphi _{1}^{\prime }\nabla ^{2}\varphi _{1}^{\prime }%
\right]  \nonumber \\
&&{\it T}\exp \left[ -i\frac{2mc}{\hslash }\sqrt{\frac{\pi G}{N-1}}\int
dtdx\left( \left[ \tilde{\varphi}(x)+\varphi _{1}(x)\right] \psi
_{1}^{\dagger }(x)\psi _{1}(x)-\tilde{\varphi}(x)g(x)\right) \right] \left|
\Phi (0)\right\rangle _{1}  \nonumber \\
&&_{1}\left\langle \Phi (0)\right| {\it T}^{-1}\exp \left[ i\frac{2mc}{%
\hslash }\sqrt{\frac{\pi G}{N-1}}\int dtdx\left( \left[ \tilde{\varphi}%
^{\prime }(x)+\varphi _{1}^{\prime }(x)\right] \psi _{1}^{\dagger }(x)\psi
_{1}(x)+\tilde{\varphi}^{\prime }(x)g^{\prime }(x)\right) \right]
\end{eqnarray}

In the equation above, the c-number factor inside the functional integration
on $g$, $g^{\prime }$ for large $N$ is the transition amplitude between the
evolved meta-states respectively in the presence of external classical mass
densities proportional to $g$ and $g^{\prime }$. Now consider that, inside
the meta-state space $S$ and for large $N$, the terms containing $g$ and $%
g^{\prime }$ in the double space integrals have vanishing mean square
deviations (as can be checked at any perturbative order), though their
commutators with the other terms are $O(N^{0})$, ($\tilde{\varphi},\tilde{%
\varphi}^{\prime }$ integrations imply $g,g^{\prime }\sim \psi ^{\dagger
}\psi $). The latter property forces us, when defining the expression
explicitly as the limit as $dt\rightarrow 0$ of time ordered products of
time evolution operators during time intervals $dt$, to keep the factors
depending on $g$ and $g^{\prime }$ in the right place, while the former one
allows for the replacement of the exponential of these terms with the
exponential of their average in the meta-state at that time instant. As a
result, if we remember that $\left\langle \psi _{\alpha }^{\dagger }(x)\psi
_{\alpha }(x)\right\rangle =\left\langle \psi _{1}^{\dagger }(x)\psi
_{1}(x)\right\rangle $, we have
\[
M(t)\mathrel{\mathop{\sim }\limits_{N\rightarrow \infty }}
\]
\begin{eqnarray}
&&\int {\it D}\left[ \tilde{\varphi}\right] {\it D}\left[ \tilde{\varphi}%
^{\prime }\right] {\it D}\left[ g\right] {\it D}\left[ g^{\prime }\right]
\int {\it D}\left[ \varphi _{1}\right] {\it D}\left[ \varphi _{1}^{\prime }%
\right] \exp \frac{ic^{2}}{2\hslash }\int dtdx\left[ -\varphi _{1}\nabla
^{2}\varphi _{1}+\varphi _{1}^{\prime }\nabla ^{2}\varphi _{1}^{\prime }%
\right]  \nonumber \\
&&{\it T}\exp \frac{-i2mc}{\hslash }\sqrt{\frac{\pi G}{N}}\int dtdx
\nonumber \\
&&\left( \left[ \tilde{\varphi}(x)+\varphi _{1}(x)\right] \psi _{1}^{\dagger
}(x)\psi _{1}(x)-g(x)\left[ \tilde{\varphi}(x)+\frac{m}{2c}\sqrt{\frac{GN}{%
\pi }}\int dy\frac{\left\langle \psi _{1}^{\dagger }(y)\psi
_{1}(y)\right\rangle }{|x-y|}\right] \right) \left| \Phi (0)\right\rangle
_{1}  \nonumber \\
&&_{1}\left\langle \Phi (0)\right| {\it T}^{-1}\exp i\frac{2mc}{\hslash }%
\sqrt{\frac{\pi G}{N}}\int dtdx  \nonumber \\
&&\left( \left[ \tilde{\varphi}^{\prime }(x)+\varphi _{1}^{\prime }(x)\right]
\psi _{1}^{\dagger }(x)\psi _{1}(x)-g^{\prime }(x)\left[ \tilde{\varphi}%
^{\prime }(x)+\frac{m}{2c}\sqrt{\frac{GN}{\pi }}\int dy\frac{\left\langle
\psi _{1}^{\dagger }(y)\psi _{1}(y)\right\rangle }{|x-y|}\right] \right) ,
\end{eqnarray}
where $\left\langle \psi _{1}^{\dagger }(y)\psi _{1}(y)\right\rangle \equiv
\left\langle \tilde{\Phi}(t)\right| \psi _{1}^{\dagger }(y,t)\psi
_{1}(y,t)\left| \tilde{\Phi}(t)\right\rangle $. By integrating over $%
g,g^{\prime },\tilde{\varphi},\tilde{\varphi}^{\prime }$, we get
\[
M(t)\mathrel{\mathop{\sim }\limits_{N\rightarrow \infty }}
\]
\begin{eqnarray*}
&&\int {\it D}\left[ \varphi _{1}\right] {\it D}\left[ \varphi _{1}^{\prime }%
\right] \exp \frac{ic^{2}}{2\hslash }\int dtdx\left[ -\varphi _{1}\nabla
^{2}\varphi _{1}+\varphi _{1}^{\prime }\nabla ^{2}\varphi _{1}^{\prime }%
\right] \\
&&{\it T}\exp \left[ \frac{iG}{\hslash }m^{2}\int dt\int dxdy\frac{\psi
^{\dagger }(x)\psi (x)\left\langle \psi ^{\dagger }(y)\psi (y)\right\rangle
}{|x-y|}-\frac{i2mc}{\hslash }\sqrt{\frac{\pi G}{N}}\int dtdx\varphi
_{1}(x)\psi ^{\dagger }(x)\psi (x)\right] \left| \Phi (0)\right\rangle \\
&&\left\langle \Phi (0)\right| {\it T}^{-1}\exp \left[ \frac{-iG}{\hslash }%
m^{2}\int dt\int dxdy\frac{\psi ^{\dagger }(x)\psi (x)\left\langle \psi
^{\dagger }(y)\psi (y)\right\rangle }{|x-y|}+\frac{i2mc}{\hslash }\sqrt{%
\frac{\pi G}{N}}\int dtdx\varphi _{1}^{\prime }(x)\psi ^{\dagger }(x)\psi (x)%
\right]
\end{eqnarray*}
omitting the by now irrelevant index in $\psi _{1}$, and finally, after
integrating out $\varphi _{1},\varphi _{1}^{\prime }$,
\[
\lim_{N\rightarrow \infty }M(t)\equiv \left| \Phi (t)\right\rangle
\left\langle \Phi (t)\right| =
\]
\begin{eqnarray}
&&{\it T}\exp \left[ \frac{iG}{\hslash }m^{2}\sum_{\alpha <\beta }\int
dt\int dxdy\frac{\psi ^{\dagger }(x,t)\psi (x,t)\left\langle \Phi (t)\right|
\psi ^{\dagger }(y,t)\psi (y,t)\left| \Phi (t)\right\rangle }{|x-y|}\right]
\left| \Phi (0)\right\rangle  \nonumber \\
&&\left\langle \Phi (0)\right| {\it T}^{-1}\exp \left[ \frac{-iG}{\hslash }%
m^{2}\sum_{\alpha <\beta }\int dt\int dxdy\frac{\psi ^{\dagger }(x,t)\psi
(x,t)\left\langle \Phi (t)\right| \psi ^{\dagger }(y,t)\psi (y,t)\left| \Phi
(t)\right\rangle }{|x-y|}\right] ,  \label{SN1}
\end{eqnarray}
where the normalization is automatically correct, as the resulting dynamics,
though nonlinear, is unitary. It should be remarked that, to make the
derivation more rigorous, the Newton potential should be replaced with \ a
regularized potential like $1/(|x-y|+\lambda )$ with $\lambda >0$, and the
Laplacian with the corresponding inverse. One should first take the limit $%
N\rightarrow \infty $, then take $dt\rightarrow 0$ and remove the
regularization, $\lambda \rightarrow 0$.

Eq. (\ref{SN1}) coincides with the time evolution of the SN model in the
interaction representation, which is then the $N\rightarrow \infty $ limit
of our model. Just as in condensed matter physics, this limit suppresses
fluctuations (quantum fluctuations here) and preserves mean field features
only. In the present case the $N\rightarrow \infty $ limit keeps the
presence of localized states, but wipes out the non unitary evolution and
then the ability of generating decoherence.

In conclusion the present construction is the first derivation of the SN
model from a well defined quantum model producing both classical
gravitational interactions, localization and decoherence. In fact the model
was usually presented, up to now, as some sort of mean field approximation
of a not yet well specified theory incorporating self-interactions.

\section{Localization: analytical and numerical results}
\label{section4}

In this Section we want to explore the possibility that even for a genuinely
isolated system an entropy growth takes place, within a non relativistic
quantum description, owing to the gravitational interaction \cite{sergio2}. In order to do
that we use the model introduced in the previous Sections \cite{sergio1,sergio3,sergio4},
whose non unitary dynamics
reproduces both the classical aspects of the gravitational interaction and
wave function localization of macroscopic bodies\cite{sergio1}.

Even in a one particle model like the Schroedinger-Newton theory, without
added unobservable degrees of freedom, one can exhibit stationary localized
states \cite{kumar}. However, apart from the price paid in abandoning the
traditional linear setting of QM, they sound quite unrealistic, as the
initial linear momentum uncertainty is expected to lead to a spreading of
the probability density in space. On the other hand a theory of wave
function localization has to keep localization during time evolution. Our
model offers a way out of this apparent paradox, leading to a spreading that
consists in the emergence of delocalized ensembles of localized pure states.

\subsection{Analytical results}

In Section \ref{section2} it has been shown also that, omitting the internal wave function, a localized
meta-state of an isolated homogeneous spherical macroscopic body of radius $R$
and mass $M$ can be represented by (see Eq. (\ref{gaussian}))
\begin{equation}
\tilde{\Psi}_{0}(X,Y)\propto \exp \frac{-\left| X-Y\right| ^{2}}{2\Lambda
^{2}}\exp \frac{-\left| X+Y\right| ^{2}}{2\Lambda ^{2}},\;\;\Lambda ^{2}=%
\frac{\hslash }{\sqrt{\alpha GM^{3}/R^{3}}},  \label{gaussian1}
\end{equation}
where $\alpha \sim 10^{0}$ and $G$ is the gravitational constant. In Eq. (%
\ref{gaussian1}) $X$ and $Y$ are respectively the position of the center of
mass of the green and the red meta-body. The first factor is proportional to
the wave function of the relative motion and, for bodies of ordinary density
$\sim 1gm/cm^{3}$ and whose mass exceeds $\sim 10^{11}$ proton masses, it
represents its ground state\cite{sergio1}. The second factor is
proportional to the wave function of the center of meta-mass $(X+Y)/2$ and
spreads in time as usual for the free motion of the center of mass of a body
of mass $2M$, so that after a time $t$, in the absence of external forces,
the meta-wave function becomes
\begin{equation}
\tilde{\Psi}_{t}(X,Y)\propto \exp \frac{-\left| X-Y\right| ^{2}}{2\Lambda
^{2}}\exp \frac{-\left| X+Y\right| ^{2}/4}{\Lambda ^{2}/2+i\hslash t/M}%
\equiv \exp \left[ -\alpha _{0}\left| X-Y\right| ^{2}\right] \exp \left[
-\alpha _{t}\left| X+Y\right| ^{2}\right] .  \label{evolvedgaussian}
\end{equation}
In order that this may be compatible with the assumption that gravity
continuously forces localization, the spreading of the physical state must
be the outcome of a growth of the corresponding entropy. This initially
vanishes, since the initial meta-wave function is unentangled:
\begin{equation}
\tilde{\Psi}_{0}(X,Y)\propto \exp \frac{-X^{2}}{\Lambda ^{2}}\exp \frac{%
-Y^{2}}{\Lambda ^{2}},
\end{equation}
and then the corresponding physical state obtained by tracing out $Y$ is
pure. If one evaluates the physical state according to
\begin{equation}
\rho _{t}(X,X^{\prime })=\int dY\tilde{\Psi}_{t}(X,Y)\tilde{\Psi}_{t}^{\ast
}(X^{\prime },Y),
\end{equation}
one finds that the space probability density is given by
\begin{equation}
\rho _{t}(X,X)=\left[ \frac{8\alpha _{0}(\alpha _{t}+\bar{\alpha}_{t})}{\pi
(\alpha _{t}+\bar{\alpha}_{t}+2\alpha _{0})}\right] ^{3/2}\exp \left[ -\frac{%
8\alpha _{0}(\alpha _{t}+\bar{\alpha}_{t})}{(\alpha _{t}+\bar{\alpha}%
_{t}+2\alpha _{0})}X^{2}\right] \propto \exp \frac{-2\Lambda ^{2}X^{2}}{%
\Lambda ^{4}+2\hslash ^{2}t^{2}/M^{2}}  \label{probability}
\end{equation}
Parenthetically it is worth while to remark that this spreading of the
probability density is slower than the one ensuing from the spreading of the
wave function in the absence of the gravitational self-interaction, which
leads to
\begin{equation}
\rho _{t}(X,X)\propto \exp \frac{-2\Lambda ^{2}X^{2}}{\Lambda ^{4}+4\hslash
^{2}t^{2}/M^{2}},  \label{traditionalprobability}
\end{equation}
and that both are extremely slow, as their typical time, for macroscopic
bodies of ordinary density, is $\sim 10^{3}\sec $ independently from the
mass, as can be checked by means of Eqs. (\ref{probability}) and (\ref
{traditionalprobability}).

If this spreading is due to entropy growth only, rather than to the usual
spreading of the wave function, the corresponding entropy $S_{t}$ is
expected to depend approximately on the ratio between the final and the
initial space volumes roughly occupied by the two gaussian densities,
according to
\begin{equation}
S_{t}\sim K_{B}\frac{3}{2}\ln \left[ \frac{\alpha _{t}+\bar{\alpha}%
_{t}+2\alpha _{0}}{2(\alpha _{t}+\bar{\alpha}_{t})}\right] ,
\label{approxentropy}
\end{equation}
at least for large enough times. (Linear momentum probability density does
not depend on time.) Of course this corresponds to approximating the mixed
state by means of an ensemble of $N$ equiprobable localized states, which is
legitimate if $N$ is large enough. In order to evaluate the entropy of the
state represented by $\rho _{t}(X,X^{\prime })$ and to check Eq. (\ref
{approxentropy}), we use the possibility, in this approximation, of linking
the entropy
\begin{equation}
S_{t}=-K_{B}\text{ }Tr\left[ \rho _{t}\ln \rho _{t}\right] =K_{B}\ln N
\label{equientropy}
\end{equation}
with the purity
\begin{equation}
Tr\left[ \rho _{t}^{2}\right] =\frac{1}{N},  \label{equipurity}
\end{equation}
where of course
\begin{equation}
\rho _{t}^{2}(X,X^{\prime })=\int dX^{\prime \prime }\rho _{t}(X,X^{\prime
\prime })\rho _{t}(X^{\prime \prime },X^{\prime }).
\end{equation}
By an explicit computation we get
\begin{equation}
Tr\left[ \rho _{t}^{2}\right] =\int dX\rho _{t}^{2}(X,X)=\frac{\left[
4\alpha _{0}(\alpha _{t}+\bar{\alpha}_{t})\right] ^{3}}{\left[ \left(
2\alpha _{t}\bar{\alpha}_{t}+6\alpha _{t}\alpha _{0}+6\bar{\alpha}_{t}\alpha
_{0}+2\alpha _{0}^{2}\right) ^{2}-4\left( \bar{\alpha}_{t}-\alpha
_{0}\right) ^{2}\left( \alpha _{t}-\alpha _{0}\right) ^{2}\right] ^{3/2}},
\end{equation}
and, for large times, namely small $\alpha _{t}$, one can keep in this
result just the leading term in $\alpha _{t}$, that is
\begin{equation}
Tr\left[ \rho _{t}^{2}\right] \sim \left( \frac{\alpha _{t}+\bar{\alpha}_{t}%
}{2\alpha _{0}}\right) ^{3/2},
\end{equation}
which, by using Eqs. (\ref{equientropy},\ref{equipurity}), gives
\begin{equation}
S_{t}\sim -K_{B}\frac{3}{2}\ln \left( \frac{\alpha _{t}+\bar{\alpha}_{t}}{%
2\alpha _{0}}\right) =K_{B}\frac{3}{2}\ln \left( \frac{\Lambda ^{4}+4\hslash
^{2}t^{2}/M^{2}}{\Lambda ^{4}}\right) ,
\end{equation}
which differs from the leading term in Eq.(\ref{approxentropy}) by an
irrelevant quantity $(3/2)K_{B}\ln 2$.

It is worth while to remark that, while the present model is expected to be
just a low energy approximation to a conceivable more general theory, its
present application to a free motion is expected basically to reproduce the
possible exact outcome of the latter. In fact the analysis refers to the
rest frame of the probability density and implies exceedingly small
velocities, as can be checked by taking the Fourier transform of the wave
function in Eq. (\ref{evolvedgaussian}).

The above analysis addresses what can be called fundamental entropy growth, which in principle could be defined
even for the universe as a whole. For real, only approximately isolated,
systems one would expect that usually, however small the coupling with the
environment may be, the corresponding entanglement entropy\cite{gemmer} is
easily large enough to overshadow the fundamental one. This is the
thermodynamical counterpart of the overshadowing of fundamental decoherence
\cite{sergio1} by the environment induced one\cite{zurek,zurek1}.

\subsection{Numerical results}

The key hypothesis behind any model of emergent classicality is that for
some reason (the surrounding environment or a fundamental noisy source)
unlocalized macroscopic states get localized, i.e. that coherences for space
points farther than some localization lengths vanish \cite{zurek}. In such a
way embarrassing quantum superpositions of distinct position states of
'macroscopic' bodies are avoided, thus recovering an essential element of
reality of our classical realm of predictability: the elementary fact that
bodies are observed to occupy quite definite positions in space. The aim of
the present subsection is to check that feature, which makes the model a viable
localization model, without using very peculiar initial conditions or any
approximation, and independently of the heuristic functional formulation \cite{sergioFil2}. In
order to make the problem numerically tractable we take a rotational
invariant initial state of an isolated ball of ordinary matter density and a
total mass just above the mass threshold, namely in the least favorable
conditions to exhibit dynamical localization \cite{sergioFil2}.

Consider now a uniform matter ball of mass $M$ and radius $R$. Within the
model the Schroedinger equation for the meta-state wave function $\Xi
(X,Y,t) $ is given by
\begin{equation}
i\hslash \frac{\partial \Xi }{\partial t}=\left[ -\frac{\hslash ^{2}}{2M}%
(\nabla _{X}^{2}+\nabla _{Y}^{2})+V(\left| X-Y\right| )\right] \Xi
\label{schroedinger1}
\end{equation}
where $X$ and $Y$ respectively denote the position of the center of mass of
the physical body and of its hidden partner, while $V$ is the (halved)
gravitational mutual potential energy of the two interpenetrating
meta-bodies, which, as can be shown by an elementary calculation, is
\begin{equation}
V(r)=\frac{1}{2}GM^{2}\left( \frac{\theta
(2R-r)(80R^{3}r^{2}-30R^{2}r^{3}+r^{5}-192R^{5})}{160R^{6}}-\frac{\theta
(r-2R)}{r}\right)  \label{potential}
\end{equation}
where $\theta $ denotes the Heaviside function. Observe that our final
result can be immediately reread as the solution for the whole set of
parameters obtained by the scaling:
\[
t\rightarrow \lambda t,\;\;\;\;\;M\rightarrow \lambda
^{-1/5}M,\;\;\;\;\;\;X\rightarrow \lambda ^{3/5}X,\;\;\;\ \;R\rightarrow
\lambda ^{3/5}R\;\;\;\;\;\;
\]
\begin{equation}
\Xi \left( X,Y;t\right) \rightarrow \Xi \left( \lambda ^{-3/5}X,\lambda
^{-3/5}Y;t\right)
\end{equation}
where $\lambda $ is a real positive dimensionless parameter. This is
consistent with the expression for $\tau _{g}$, the latter giving $\tau
_{g}\rightarrow \lambda \tau _{g}$. Besides, for consistency, we note that
the mass cannot cross the threshold, as the latter scales with the same
power law of the mass itself $M_{t}\rightarrow \lambda ^{-1/5}M_{t}$.

If we separate Eq. (\ref{schroedinger1}) into the equation for the relative
motion and that for the center of mass, then for ordinary matter density $\varrho \sim 10^{24}m_{p}/\mathop{\rm cm}^{3}$ and $M$
above the threshold $\sim 10^{11}m_{p}$, the former admits
bound meta-states of width $\Lambda _{G}\sim (m_{p}/M)^{1/2}cm$,
corresponding to small oscillations around the minimum of the gravitational
potential. In particular an untangled localized meta-state, corresponding to
a physical pure state is:
\begin{equation}
\Psi _{TOT}=\Psi _{0}(X)\Psi _{0}(Y)=\Psi _{0}(\left[ X+Y\right] /2)\Psi
_{0}(\left[ X-Y\right] /2),  \label{pure}
\end{equation}
where
\[
\Psi _{0}(\left[ X-Y\right] /2)=\left( \Lambda _{G}^{2}\pi \right)
^{-3/4}\exp [-\left| X-Y\right| ^{2}/(2\Lambda _{G}^{2})];\;\ \ \;\Lambda
_{G}=(8\hslash ^{2}R^{3}/GM^{3})^{1/4}
\]
is proportional to the ground meta-state of the relative motion in the
hypothesis $\Lambda _{G}\ll R$.

If $\Xi (X,Y)\equiv \psi (\left[ X+Y\right] /2)\phi (X-Y)$, the equation for
$\phi $, due to the spherical symmetry, reduces to the radial equation for $%
\chi (r)\equiv \phi (\left| X-Y\right| )$. This equation has been solved
numerically by the algorithm obtained from the space discretization over $%
10^{4}$ points of the equation (Fig.1)
\begin{equation}
\left[ 1+\frac{1}{2}iHdt/\hslash \right] u(r,t+dt)=\left[ 1-\frac{1}{2}%
iHdt/\hslash \right] u(r,t)+o(dt^{2})
\end{equation}
in the interval $(-R/2,R/2)$, where $H$ denotes the radial Hamiltonian for $%
u(r,t)\equiv r\chi (r,t)$.$\;$Such a procedure assures the stability of the
state-vector norm during the time evolution and is second-order accurate\cite
{recipes}.

\begin{figure}[tbph]
\centerline{\includegraphics[scale=0.4]{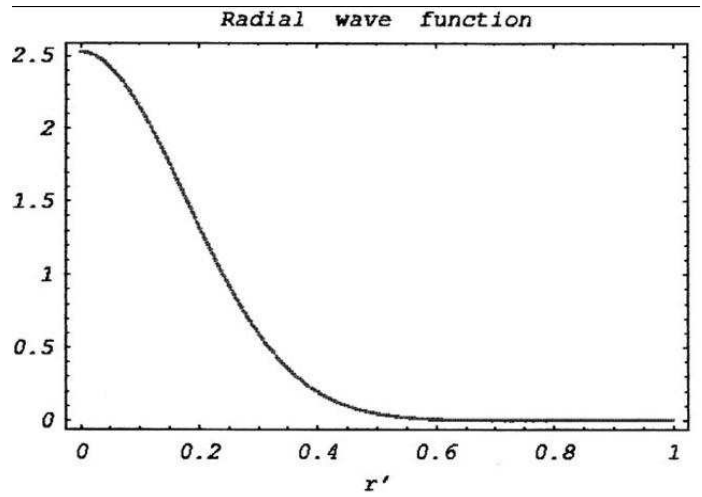}}
\vspace*{8pt}
\caption{Initial radial wave function $\chi \left( r^{\prime }\right) $, with
$r^{\prime }\equiv $ $r/R$, for the internal state of the two meta-bodies.}
\label{fig:fig1}
\end{figure}

A uniform ball of mass $M=0.38\times 10^{12}m_{p}$ and radius $R=4.8\times
10^{-5}\mathop{\rm cm}$ was considered, with initial conditions like in Eq.(\ref{pure}), but
for $\Lambda _{G}\sim 1.6\times 10^{-6}\mathop{\rm cm}$ replaced by $\Lambda =5.6\Lambda _{G}$.
The solution of Eq. (\ref{schroedinger1}) is then obtained as the product of $\phi $ and the
analytical solution
\begin{equation}
\psi (\left[ X+Y\right] /2,t)\propto \exp \left[ \frac{-\left| X+Y\right|
^{2}/4}{\Lambda ^{2}/2+i\hslash t/M}\right]
\end{equation}
of the center of meta-mass equation (Fig. 2).

\begin{figure}[tbph]
\centerline{\includegraphics[scale=0.4]{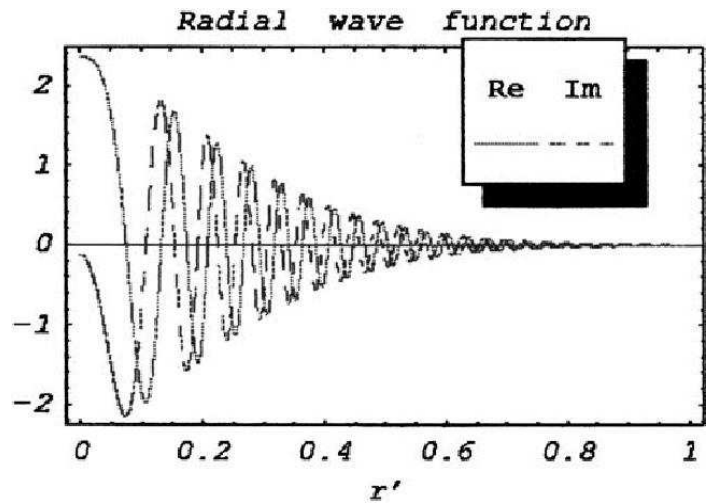}}
\vspace*{8pt}
\caption{Radial wave function $\chi \left( r^{\prime }\right) $, with $r^{\prime }\equiv $ $r/R$, for the internal state of the two meta-bodies at
the final simulation time $t=10\mathop{\rm s}$.}
\label{fig:fig2}
\end{figure}

The physical state $\rho $ is evaluated by tracing out the hidden body
\begin{equation}
\rho (X;X^{^{\prime }})=\int dY\Xi (X,Y)\Xi ^{*}(X^{\prime },Y).
\label{phstate}
\end{equation}

After an evolution time $t=10\mathop{\rm s}$ ($\sim \tau _{g}$) the
function $\widetilde{\varrho }\left(X_{1},X_{1}^{\prime }\right) \equiv \rho (X_{1},0,0;X_{1}^{^{\prime }},0,0)$
can be represented as in Fig. 4. To compare with the free evolution, in
which the dynamics gives the usual spreading, we have also shown the
corresponding function in Fig. 3.
\begin{figure}[tbph]
\centerline{\includegraphics[scale=0.4]{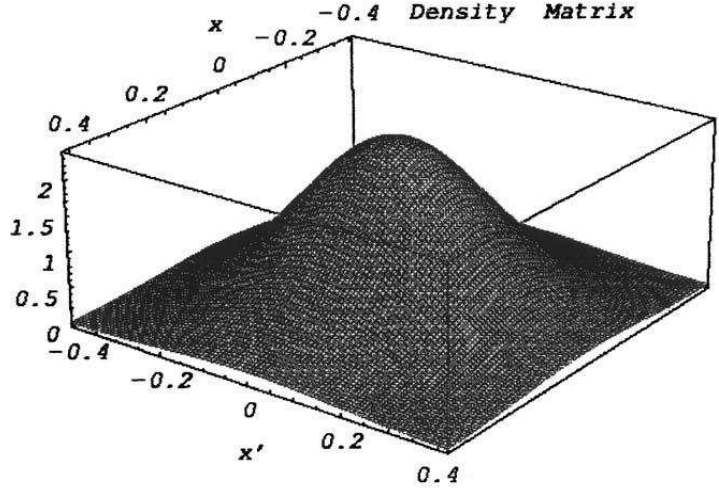}}
\vspace*{8pt}
\caption{Modulus of the reduced density matrix $\left| \rho \left(X,X^{\prime }\right) \right| $ in the absence of gravity in the
plane $X_{1},X_{1}^{\prime }$ with $X_{2}=X_{2}^{\prime}=X_{3}=X_{3}^{\prime }=0$,$\;$at the evolution time $t=10$ $\mathop{\rm s}$.}
\label{fig:fig3}
\end{figure}
\begin{figure}[tbph]
\centerline{\includegraphics[scale=0.4]{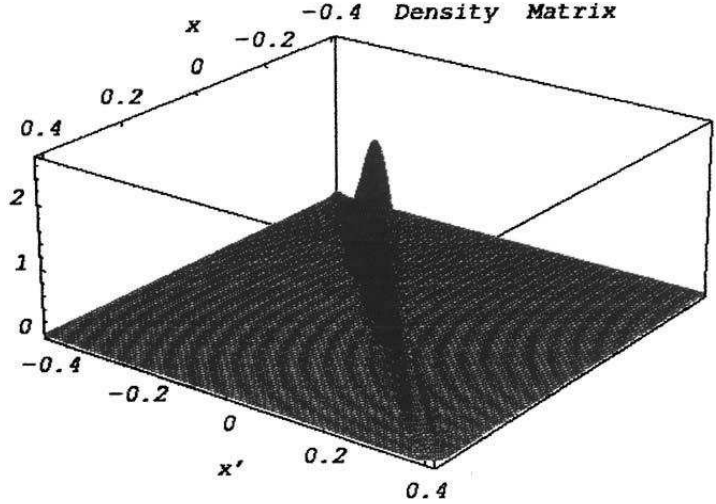}}
\vspace*{8pt}
\caption{ Modulus of the reduced density matrix $\left| \rho \left(X,X^{\prime }\right) \right| \;$with the inclusion of gravity in the plane $X_{1},X_{1}^{\prime }$ with $X_{2}=X_{2}^{\prime }=X_{3}=X_{3}^{\prime }=0\;$ at the evolution time $t=10$ $\mathop{\rm s}$.}
\label{fig:fig4}
\end{figure}

The final result can even be fitted by the product of two Gaussian
functions:
\begin{equation}
\widetilde{\varrho }\left( X_{1},X_{1}^{\prime }\right) =\exp
[-(X_{1}+X_{1}^{^{\prime }})^{2}/\Lambda _{+}^{2}]\exp
[-(X_{1}-X_{1}^{^{\prime }})^{2}/\Lambda _{-}^{2}]  \label{state}
\end{equation}
where $\Lambda _{+}=0.27R\sim 1.3\times 10^{-5}\mathop{\rm cm}$ and $\Lambda _{-}=1.8\times 10^{-2}R\sim 8.1\times 10^{-7}\mathop{\rm cm}$, while the free evolution, ignoring the gravitational (self-)interaction,
would give the same product structure with $\Lambda _{+}=\Lambda _{-}\sim
1.3\times 10^{-5}\mathop{\rm cm}\;$ (see Figs. 5 and 6). In spite of the fact that the fit has been performed
by simple Gaussian functions, with the height and the size as independent
parameters, the result turns out to be very accurate.
\begin{figure}[tbph]
\centerline{\includegraphics[scale=0.4]{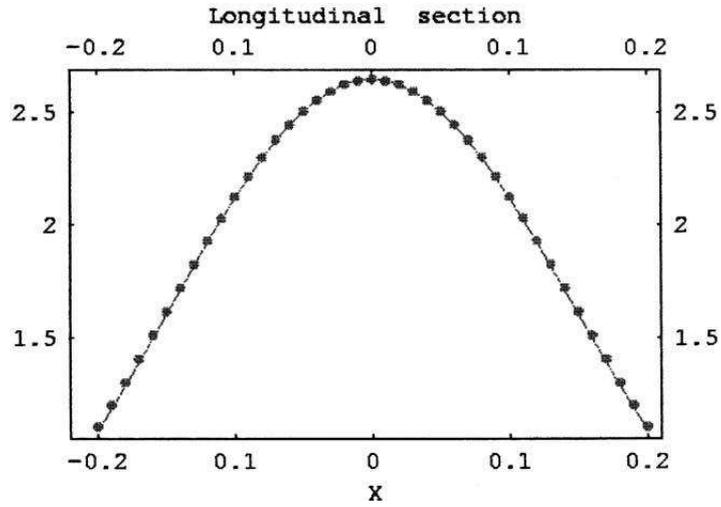}}
\vspace*{8pt}
\caption{A longitudinal Gaussian profile corresponding to the function $\left| \widetilde{\rho }\left( X,X\right) \right| $ has been superimposed to
the array of points obtained from the numerical simulation. Lengths $X$ on the x-axis are measured in units of the radius $R$.}
\label{fig:fig5}
\end{figure}
\begin{figure}[tbph]
\centerline{\includegraphics[scale=0.4]{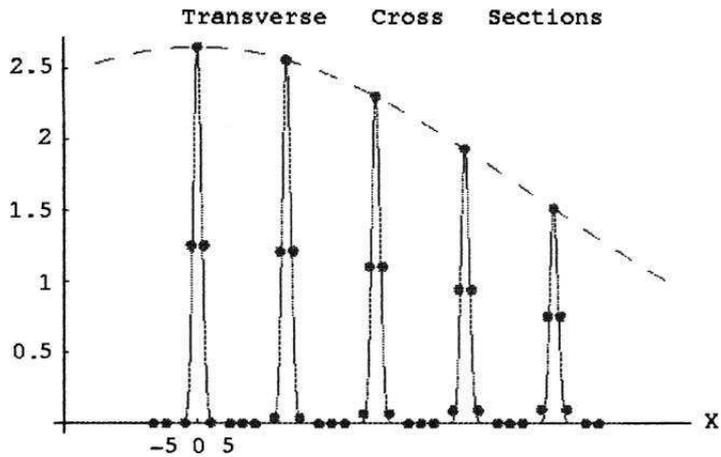}}
\vspace*{8pt}
\caption{Gaussian transverse cross-sections corresponding to the function $\left| \widetilde{\rho }\left( X,-X+k\ast 0.08R\right) \right| ,$ $k=0,1,2,3,4$ have been superimposed to the arrays of points obtained from the numerical simulation. The dashed line represents the longitudinal cross
section. Lengths $X$ on the x-axis are measured in units of $10^{-2}R$.}
\label{fig:fig6}
\end{figure}

It should be remarked that the exact rotational invariance of the initial
state, which rather artificially leads to a degenerate final state, does not
play any special role. In fact, due to the unitarity in the enlarged Hilbert
space, modifying slightly the initial state would result in a slight
modification of the final meta-state, and ultimately of the physical state.
Nevertheless the choice of a pure initial state corresponds to a precise
measurement, which represents in our case a rather ideal situation. On the
other hand the simplicity of the state constraint renders even conceptually
more precise such a measurement with respect to the case of a system
interacting with a complex environment, where the state of the environment
after the measurement is in general a complex functional of the system
state, to a large extent uncontrollable. In any case the present calculation
has to be considered only as a first step in the analysis of an intrinsic
non-unitary model.

To estimate the final entropy production, consider that $\delta
(X_{2})\delta (X_{3})\widetilde{\varrho }(X_{1},X_{1}^{\prime })\delta
(X_{2}^{\prime })\delta (X_{3}^{\prime })$ represents the state resulting
from a measurement on $\rho $, giving $X_{2}=X_{3}=$ $0$, by which one means
that the uncertainties are smaller than the typical scale of variation of $%
\rho $. The entropy of $\widetilde{\varrho }$ then gives a lower bound for
the entropy of $\rho $. To estimate the entropy $S\left[ \widetilde{\varrho }%
\right] $ of $\widetilde{\varrho }$, we evaluate its purity
\begin{equation}
Tr\widetilde{\varrho }^{2}=\int dX_{1}dX_{1}^{\prime }\widetilde{\varrho }%
\left( X_{1},X_{1}^{\prime }\right) \widetilde{\varrho }\left( X_{1}^{\prime
},X_{1}\right) ,
\end{equation}
which, for $\widetilde{\varrho }$ replaced by its analytical fit (Eq. \ref
{state}), gives $Tr\widetilde{\varrho }^{2}=\Lambda _{-}/\Lambda _{+}\sim
6.\times 10^{-2}$. If, for simplicity, we consider the corresponding
ensemble as one of $N$ equiprobable states, then
\begin{equation}
Tr\widetilde{\varrho }^{2}=\sum_{j=1}^{N}\frac{1}{N^{2}}=\frac{1}{N}%
\Rightarrow N\sim 17;\;S\left[ \widetilde{\varrho }\right] \sim K_{B}\log 17.
\end{equation}
If we approximate $\rho \,$as the direct product of three equivalent $%
\widetilde{\varrho }$, namely we omit the entanglement between the three
Cartesian coordinates, which is absent in the initial pure state and would
stay so if the potential in Eq. (\ref{potential}) were replaced by its
quadratic approximation, then the total entropy is $S\left[ \rho \right] =3S%
\left[ \widetilde{\varrho }\right] $, corresponding to $N^{3}$ equiprobable
states.

Independently of any approximation $\rho (X;X^{^{\prime }})$, as the kernel
of a compact positive semi-definite Hermitian operator of unit trace \cite
{vNeumann}, can be diagonalized as
\begin{equation}
\rho (X;X^{^{\prime }})=\sum_{j}p_{j}\psi _{j}(X)\psi _{j}^{\ast }(X^{\prime
});\;p_{j}\geq 0;\;\sum_{j}p_{j}=1;\;\left\langle \psi _{j}|\psi
_{k}\right\rangle =\delta _{jk},
\end{equation}
where the above approximate estimate makes us expect that $%
-\sum_{j}p_{j}\log p_{j}\sim 3\log 17$. This result has to be compared with
a value $\sim 3\log \left[ \Lambda _{+}/\Lambda _{G}\right] \sim 3\log 8$
corresponding to a naive counting where the orthogonal states $\psi _{j}$
above are assumed localized and approximately non overlapping. This small
discrepancy is not surprising, as these states are expected to include
contributions from several low lying bound meta-states, so that they are
orthogonal in spite of the overlapping of their probability densities, due
to their space oscillations.

The qualitative agreement between our two estimates of entropy, the one
corresponding to the computed density matrix and the other corresponding to
the approximation of the mixed state by means of an ensemble of equiprobable
localized states which occupy the volume roughly occupied by the density $%
\rho (X;X)$, makes it natural to assume that the states $\psi _{j}$
diagonalizing $\rho $ are localized. To be more precise, this would be
strictly true, without ambiguities, after breaking the exact rotational
invariance, which, as mentioned, is expected to introduce artificial
degeneracies in the density operator.

The present result strengthens our confidence in the most relevant
peculiarities of the localization phenomenology ensuing from the model,
which make it {\it in principle} distinguishable both from the other
proposed models \cite{ghirardi1,pearle2} and possibly from the competing action
of the environment-induced decoherence \cite{zurek}. In particular the model
presents a sharp threshold, below which localization is practically absent,
and a localization time $\tau _{g}\propto M^{-5/3}$ rapidly decreasing, as
the mass is increased. Furthermore the threshold mass $M_{t}\propto \rho
^{1/10}$ is remarkably robust with respect to mass density variations.

\section{Entropy growth and thermalization: first numerical results}
\label{section5}

In this Section we show the ability of De Filippo's model to reproduce a gravity-induced relaxation towards thermodynamic
equilibrium even for a closed system. We present results of numerical simulations on two systems with increasing complexity: two particles in an harmonic trap interacting via an `electrical' delta-like potential and gravitational interaction \cite{entropy1}, and an harmonic nanocrystal within a cubic geometry \cite{entropy2}. These preliminary studies allow us to perform a first step towards the simulation of macroscopic systems where the Second Law of thermodynamics is more relevant.

\subsection{Two-particle system}

In this Section we summarize the results of a simulation, carried out on a simple system of two interacting particles \cite{entropy1}. This has been a first but necessary step in demonstrating the ability of the De Filippo's model to reproduce a gravity-induced relaxation towards thermodynamic
equilibrium even for a closed system \cite{entropy1}.

More specifically we consider the two particles in an harmonic trap,
interacting with each other through `electrostatic' and gravitational
interaction, whose `physical' Hamiltonian, in the ordinary
(first-quantization) setting, is
\begin{equation}
H_{ph}\left( \mathbf{x}_{1},\mathbf{x}_{2}\right) =
\sum\limits_{i=1}^{2}\biggl(-\frac{\hbar ^{2}}{2\mu }\Delta _{\mathbf{x}%
_{i}}+\frac{\mu}{2} \omega ^{2}\mathbf{x}_{i}^{2}\biggr)+\sum\limits_{i<j=1}^{2}\left(\frac{4\pi \hbar ^{2}l_{s}}{%
\mu }\delta (\mathbf{x}_{i}-\mathbf{x}%
_{j})-\frac{G\mu ^{2}}{\left\vert \mathbf{x}_{i}-\mathbf{x}%
_{j}\right\vert }\right);  \label{mol1}
\end{equation}
here $\mu $ is the mass of the particles, $\omega $ is the frequency of the
trap and $l_{s}$ is the s-wave scattering length. We are considering a
`dilute' system, such that the electrical interaction can be assumed to have
a contact form with a dominant s-wave scattering channel. We take numerical
parameters that make the electrical interaction at most comparable
with the oscillator's energy, while gravity enters the problem as an higher
order correction. The model, in
the (first-quantization) ordinary setting, is defined by the following general meta-Hamiltonian:
\begin{equation}
H_{TOT}=H_{ph}\left( \mathbf{x}_{1},\mathbf{x}_{2}\right) +H_{hid}\left(
\widetilde{\mathbf{x}}_{1},\widetilde{\mathbf{x}}_{2}\right) +H_{int}\left(
\mathbf{x}_{1},\mathbf{x}_{2};\widetilde{\mathbf{x}}_{1},\widetilde{\mathbf{x%
}}_{2}\right) ,
\end{equation}
with $H_{int}=G\mu^2\sum_{i<j}\biggl(\frac{1}{2\vert \mathbf{x}_i-\mathbf{x}_j\vert}+\frac{1}{%
2\vert \widetilde{\mathbf{x}}_i-\widetilde{\mathbf{x}}_j\vert}\biggr)-G\mu^2\sum_{i,j}\frac{1}{\vert \mathbf{x}_i-\widetilde{%
\mathbf{x}}_j\vert}$.

Then the time dependent physical density matrix is computed by tracing out the hidden
degrees of freedom and the corresponding von Neumann entropy is derived as
the entanglement entropy with such hidden degrees of freedom. This is obtained via a numerical simulation, by choosing as initial condition an eigenstate of the physical Hamiltonian. As a result, we find that entropy fluctuations take place, owing to
the (non unitary part of) gravitational interactions, with the initial pure
state evolving into a mixture \cite{entropy1}. The behavior of one- and two-particle von Neumann entropy as a function of time is depicted in Fig. 7 for an initial state $\vert\phi _{2}\rangle$, equal to the eigenstate of the physical energy associated with the 2nd highest energy eigenvalue $E_2$ of the two-particle system under study. The following values of the physical parameters $l_{s}$, $\mu $
and $\omega $ have been chosen: $l_{s}=5.5\cdot 10^{-8}m$, $\mu =1.2\cdot 10^{-24}kg$ and $%
\omega =4\pi \cdot 10^{3}s^{-1}$ (which are compatible with current
experiments with trapped ultracold atoms \cite{exp1} and complex molecules
\cite{exp2}) together with an artificially augmented 'gravitational constant' $%
G=6.67408\times 10^{-6}m^{3}kg^{-1}s^{-2}$ ($10^{5}$ times the real
constant).
\begin{figure}[tbph]
\centerline{\includegraphics[scale=0.8]{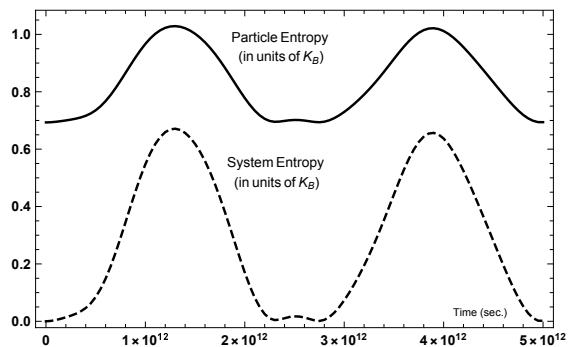}}
\vspace*{8pt}
\caption{ Time evolution of one-particle and two-particle entropies (respectively, curve above and below) for the initial state chosen.}
\label{fig:fig7}
\end{figure}

Due to gravity-induced fluctuations the system
entropy shows a net variation over very long times, at variance with the
case without gravitational interaction with the hidden
system, in which it would have been a constant of motion. At the same
time, single particle entropy, which in the ordinary setting is itself
constant, shows now a similar time modulation.
It was to be expected that for a microscopic system the recurrence times are relatively short, due to the low dimensionality of the Hilbert meta-space corresponding to the microcanonical ensemble; on the contrary the recurrence times of macroscopic systems may be longer than the age of the universe. Actually the recurrence is even more irrelevant since no system can be considered closed for long times.
The expectation of physical energy has been verified to be a constant as well,
meaning that the non unitary term has no net energy associated with itself,
but is purely fluctuational.

\subsection{Nanocrystal}

In this Subsection we switch to a bit more complex model of a
three-dimensional system, that is still computationally tractable. Taking
the simple model of an harmonic nanocrystal, in which we consider as in the
previous case an artificially augmented gravitational constant, turns out to
be a viable choice: indeed the result is a simple and nice formula for thermalization that
can be efficiently simulated numerically  \cite{entropy2}.

We consider a cubic crystal of volume $V=L^{3}$. Each phonon propagating
within the crystal is supposed to fill it homogeneously, and its
gravitational mass is given by $\hbar \omega /c^{2}$, where $\omega $ is the
phonon's angular frequency. The total Hamiltonian is given by
\begin{equation}
H_{G}=H_{ph}+H_{hid}+H_{int}
\end{equation}
where
\begin{eqnarray}
H_{ph}&=&\sum_{\mathbf{k}s}\hbar \omega _{\mathbf{k}s}\widehat{n}_{\mathbf{k}s}-\frac{%
1}{4}C(G,V)\sum_{\mathbf{k}s}\sum_{\mathbf{k}^{\prime }s^{\prime }}\left(
\frac{\hbar \omega _{\mathbf{k}s}}{c^{2}}\right) \left( \frac{\hbar \omega _{%
\mathbf{k}^{\prime }s^{\prime }}}{c^{2}}\right) \widehat{n}_{\mathbf{ks}}%
\widehat{n}_{\mathbf{k}^{\prime }s^{\prime }}, \\
H_{hid}&=&\sum_{\widetilde{\mathbf{k}}\widetilde{s}}\hbar \omega _{\widetilde{%
\mathbf{k}}\widetilde{s}}\widehat{n}_{\widetilde{\mathbf{k}}\widetilde{s}}-%
\frac{1}{4}C(G,V)\sum_{\widetilde{\mathbf{k}}\widetilde{s}}\sum_{\widetilde{%
\mathbf{k}}^{\prime }\widetilde{s}^{\prime }}\left( \frac{\hbar \omega _{%
\widetilde{\mathbf{k}}\widetilde{s}}}{c^{2}}\right) \left( \frac{\hbar
\omega _{\widetilde{\mathbf{k}}^{\prime }\widetilde{s}^{\prime }}}{c^{2}}%
\right) \widehat{\widetilde{n}}_{\widetilde{\mathbf{k}}\widetilde{s}}%
\widehat{\widetilde{n}}_{\widetilde{\mathbf{k}}^{\prime }\widetilde{s}%
^{\prime }} ,\\
H_{int}&=&-\frac{1}{2}C(G,V)\sum_{\mathbf{k}s}\sum_{\widetilde{\mathbf{k}}\widetilde{%
s}}\left( \frac{\hbar \omega _{\mathbf{k}s}}{c^{2}}\right) \left( \frac{%
\hbar \omega _{\widetilde{\mathbf{k}}\widetilde{s}}}{c^{2}}\right) \widehat{n%
}_{\mathbf{ks}}\widehat{\widetilde{n}}_{\widetilde{\mathbf{k}}\widetilde{s}%
}\ \ .\ \
\end{eqnarray}
Here $\widehat{n}$ and $\widehat{\widetilde{n}}$ are,
respectively, the physical and hidden phonon number operators. Wave numbers
are given by
\begin{equation*}
k_{i}=\frac{2\pi n_{i}}{L},\ \ \text{\ with}\ \ n_{i}=0,\pm 1,\pm 2,...\ \ \
\text{and}\ \ -\frac{\pi }{a}<k_{i}\leq \frac{\pi }{a}\ \ \ \ \ \text{(first
Brillouin zone),}
\end{equation*}
while the gravitational factor $C(G,V)$, linearly depending on $G$, is
calculated in Appendix \ref{coeffCalc}.

We assume a simple dispersion relation, corresponding to a simple
cubic crystal structure in which only the first $6$ neighbors interaction is
taken into account:
\begin{equation}
\omega _{\mathbf{k}s}=\sqrt{\frac{4K}{m}}\left\vert \sin \left( \frac{ak_{s}%
}{2}\right) \right\vert ,\ \ \ \ \ \ \ \ \ \ \ \ \ s=1,2,3
\end{equation}
where $m$ is the atomic mass, $K$ is the elastic constant and $a$ is the
lattice constant.

Indicating by $\left\vert \mathbf{n}\right\rangle =\left\vert n_{\mathbf{k}%
_{1}s_{1}}n_{\mathbf{k}_{2}s_{2}}...\right\rangle $ the state number in the
physical Fock space and by $\left\vert \widetilde{\mathbf{n}}\right\rangle
=\left\vert \widetilde{n}_{\mathbf{k}_{1}s_{1}}\widetilde{n}_{\mathbf{k}%
_{2}s_{2}}...\right\rangle $ the state number in the hidden Fock space, we
note that two generic state numbers $\left\vert \mathbf{n}^{i}\right\rangle $
and $\left\vert \widetilde{\mathbf{n}}^{j}\right\rangle $ are respectively
eigenstates of $H_{ph}$ and $H_{hid}$. This follows from the simple
observation that these Hamiltonians depend only on their respective number
operators. Let's call $E_{0,i}$ and $E_{0,j}$ their respective eigenvalues.
Now, the product $\left\vert \mathbf{n}^{i}\right\rangle \otimes \left\vert
\widetilde{\mathbf{n}}^{j}\right\rangle $ is an eigenstate of the total
Hamiltonian $H_{G}$ with eigenvalue $E_{0,i}+$ $E_{0,j}+E_{Int,i,j}$. As for
a really macroscopic (or mesoscopic) body, thermodynamic variables like energy can be defined only at a macroscopic level \cite{th2}. In particular, a thermodynamic state with
internal energy $E$ amounts, at the microscopic level, to specifying energy
within an uncertainty $\Delta E$, which includes a huge number of energy
levels of the body. For this reason, assuming an initial pure physical state
of the form
\begin{equation}
\left\vert \psi \left( 0\right) \right\rangle =\sum_{i}\gamma _{i}\left\vert
\mathbf{n}^{i}\right\rangle ,  \label{RandomState}
\end{equation}
where $\left\vert \mathbf{n}^{i}\right\rangle $ are the (physical)
energy eigenstates with eigenvalues $E_{0,i}$ close to $E$ (within $\Delta E$), the corresponding meta-state is:
\begin{equation}
\Vert \Psi \left( 0\right) \rangle \rangle =\left\vert \psi \left( 0\right)
\right\rangle \otimes \left\vert \widetilde{\psi }\left( 0\right)
\right\rangle .
\end{equation}
The state (\ref{RandomState}) is intended to be drawn uniformly at random from
the high dimensional subspace corresponding to the energy interval
considered. This is reminiscent of the notion of \textit{tipicality} introduced
in the context of the Eigenstate Thermalization Hypothesis (ETH) \cite{Deutsch,Rigol}%
, where this concept is more precisely stated by saying that state vector
above is distributed according to the Haar measure over the considered
subspace \cite{Gogolin}. To the purpose of our numerical implementation, the
simple algorithm described in Ref. \cite{Maziero} is used.

At time $t$, assuming a complete isolation of the body, we get
\begin{equation}
\Vert \Psi \left( t\right) \rangle \rangle =\sum\limits_{i,j}\gamma
_{i}\gamma _{j}e^{-\left( i/\hbar \right) \left[ E_{0,i}+E_{0,j}+E_{Int,i,j}%
\right] \ t}\left\vert \mathbf{n}^{i}\right\rangle \otimes \left\vert \widetilde{\mathbf{n}}%
^{j}\right\rangle \equiv \Gamma _{i,j}\left( t\right) \left\vert
\mathbf{n}^{i}\right\rangle \otimes \left\vert \widetilde{\mathbf{n}}^{j}\right\rangle .
\end{equation}
The physical state $\rho _{ph}$ is then given by
\begin{equation}
\rho _{ph}\left( t\right) =\sum\limits_{i,i^{\prime }}f_{i,i^{\prime
}}\left( t\right) \left\vert \mathbf{n}^{i}\right\rangle \left\langle
\mathbf{n}^{i^{\prime }}\right\vert ,
\end{equation}
with
\begin{eqnarray}
f_{i,i^{\prime }}\left( t\right) &=&\sum\limits_{j}\Gamma _{i,j}^{\ast }\left(
t\right) \Gamma _{i^{\prime },j}\left( t\right)=\gamma _{i}^{\ast }\gamma _{i^{\prime }}e^{-(i/\hbar )\left[ E_{0,i^{\prime
}}-E_{0,i}\right] t} \sum\limits_{j}\left\vert \gamma _{j}\right\vert^{2} \nonumber \\
&\times & \exp \left\{  \frac{i}{2\hbar}  C(G,V)\sum\limits_{\mathbf{k}s,%
\mathbf{k}^{\prime }s^{\prime }}\left( \frac{\hbar \omega _{\mathbf{k}s}}{%
c^{2}}\right) \left( \frac{\hbar \omega _{\mathbf{k}^{\prime }s^{\prime }}}{%
c^{2}}\right) \left[ n_{\mathbf{k}s}^{i^{\prime }}-n_{\mathbf{k}s^{\prime
}}^{i}\right] n_{\mathbf{k}^{\prime }s^{\prime }}^{j}t\right\} .
\label{coeffi}
\end{eqnarray}
This last formula is our central result. In fact the term within the
square brackets is the one responsible for the rapid phase cancelation and
diagonalization of $\rho _{ph}\left( t\right) $ in the energy basis, as we
show in the numerical simulation that follows. Incidentally, we note the
strict resemblance of Eq. (\ref{coeffi}) with Eq. (51) of Ref. \cite{sergio6}%
, expressing the phases cancelation leading to the dynamical
self-localization of a lump. This fact reflects the deep connection between the quantum measurement problem and the law
of entropy increase, as pointed out in Ref. \cite{th2}.

In order to perform a simple and viable numerical simulation of the time
evolution of the system through the explicit computation of all the terms in Eq. (%
\ref{coeffi}), we consider a nanocrystal of $10^{3}$atoms, with
the following values for the parameters: $m=3.48\times 10^{-25}kg$ (i.e. the mass of $^{210}Po$, the only chemical
element presenting a simple cubic crystal structure), $a=335\ pm$ ($L=9a$) and
$K=23.091\ N/m$. We put a huge factor $F=10^{46}$ in front of $G$ in
order to simulate the effect of gravity in a really macroscopic system
(otherwise the characteristic time of gravitational thermalization for a
system of only $10^{3}$ atoms would be much greater than the age of the
Universe!). Besides we choose $E=1.89\ \times 10^{6}\ \hbar \sqrt{K/m}\ $and
$\Delta E$ $=0.0024\ E$.

\begin{figure}[tbph]
	\centerline{\includegraphics[scale=0.4]{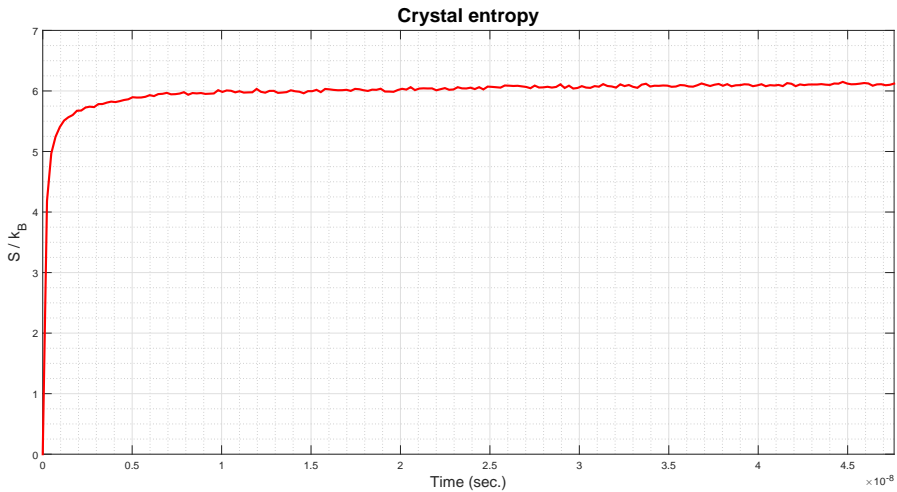}}
    \vspace*{8pt}
	\caption{Entropy as a function of time for an initial superposition of energy eigenstates within
the interval $\Delta E$, showing a monotonic increase and
		stabilization at late times, as expected for a thermalizing system.}
	\label{figure8}
\end{figure}

The numerical calculation of von Neumann entropy amounts to the repeated
diagonalization, on a discretized time axis, of the numerical matrix $%
f_{i,i^{\prime }}$. Denoting with $\lambda _{j}$ the eigenvalues of this latter
matrix at a given time, von Neumann entropy is readily computed as $%
S/k_{B}=-\sum_{j}\lambda _{j}\ln \lambda _{j}$. Its time behavior is
shown in Fig. \ref{figure8}. We can see that the gravitational term at
work reproduces correctly the expected behavior of a thermalizing system.
It is expected that the final value of entropy is the maximum value
attainable at the given internal energy $E$, provided that the state $\rho _{ph}$
contains all the available energy eigenstates (given the supposed \textit{%
typicality} of the initial state). Since the off-diagonal terms of $%
\rho _{phys}$ quickly die out in the basis of the physical energy,
consistency with the micro-canonical ensemble, and then with Thermodynamics,
is ensured. To be more precise, as it can be immediately
seen from Eq. (\ref{coeffi}), coherences still survive within the degenerate subspaces
of energy associated, in the case under study, to a permutation
symmetry of the branches $\alpha $. This amounts to a erroneous factor $3!$ in
the counting of states, which is practically irrelevant in the computation
of entropy due to the huge number of states involved. The characteristic time
for entropy stabilization depends of course on the factor $F$ multiplying $G$%
, that we have inserted to mimic the effect of a really macroscopic crystal.
Reducing $F$ amounts to an increase of this time, being the two parameters
inversely proportional, thanks to the time-energy uncertainty relation.

Up to now we have tacitly assumed in our model that the physical and hidden
crystals are perfectly superimposed on each other. This circumstance holds
true only if a strict CM\ self-localization is verified, i.e. if our crystal
mass is well above the gravitational localization threshold. In the case
under study, the big factor $F$ multiplying $G$ has to be taken into
account. Localization length can be estimated by using $\Lambda \sim \left(
\hbar \sqrt{V\ /\ FGM^{3}}\right) ^{1/2}\sim 10^{-15}m<<L\sim 10^{-9}m$.
This latter condition ensures the physical consistency of our calculation.

Finally, let us stress that the present model should be considered
just as a toy-model of a real crystal. In fact, in a real crystal the
anharmonic corrections play an important role in subsystems' thermalization,
and are of lower order with respect to (no nunitary) gravity, this latter
being qualitatively different because of its non unitary nature. Indeed it is just the non unitarity which gives rise
to the possibility of a net entropy growth for the system as a whole, so
allowing for a microscopic derivation of the Second Law of
thermodynamics. While, of course, the general model need to be tested
against properly designed future experiments, anyway we can say that it is
the first self-consistent low-energy gravity model implying in a natural way the
emergence of Thermodynamics even in a closed system.
Incidentally, within the framework of ETH \cite{Deutsch,Rigol}, a physical entropy can only be
introduced by previously applying an ad-hoc \textit{de-phasing map} to the
state of the system.

\section{Non unitary HD gravity and regularization of gravitational collapse singularities}
\label{section6}

Although higher derivative (HD) gravity has long been popular as a natural
generalization of Einstein gravity\cite{dewitt,stelle,nojiri}, since
''perturbation theory for gravity ... requires higher derivatives in the
free action''\cite{hawkinghertog}, already on the classical level it is
unstable due to negative energy fields giving rise to runaway solutions\cite
{hawkinghertog}. On the quantum level an optimistic conclusion as to
unitarity is that ''the S-matrix will be nearly unitary\cite{dewitt}''\cite
{hawkinghertog}.

A way out of the so called information loss paradox\cite{hawking2,preskill}
of black hole physics\cite{hawking1} may be precisely a fundamental
non-unitarity\cite{ellis,banks,unruh2,kay,wald}: ''For almost any initial
quantum state, one would expect ... a non vanishing probability for evolution
from pure states to mixed states''\cite{unruh2}. Though such an evolution is
incompatible with a cherished principle of quantum theory, the crucial issue
is to see if it necessarily gives rise to a loss of coherence or to
violations of energy-momentum conservation so large as to be incompatible
with ordinary laboratory physics \cite{ellis,banks,unruh2,wald}, as guessed
for Markovian effective evolution laws \cite{ellis,banks}. However one
expects that a law modeling black hole formation and evaporation, far from
being local in time, should retain a long term ``memory''\cite{unruh2,wald}.

Here a specific non-unitary realization of HD gravity is shown to be
classically stable, as well as compatible with the wavelike properties of
microscopic particles and with the assumption of a gravity-induced emergence
of classicality\cite{karolyhazy1,hawking3,diosi,ghirardi2,penrose,squires1,ellis2,anandan}.
Moreover it leads to the reading of the thermodynamical entropy of a closed
system as von Neumann entropy, or equivalently as entanglement entropy with
hidden degrees of freedom\cite{unruh2,wald}, which allows, in principle, to
overcome the dualistic nature of the notions of ordinary and
Bekenstein-Hawking (B-H) entropy \cite{bekenstein}. To be specific, the B-H
entropy\cite{bekenstein} may be identified with the von Neumann entropy of
the collapsed matter, or equivalently with the entanglement entropy between
matter and hidden degrees of freedom, both close to the smoothed
singularity. In fact the model seems to give clues for the elimination of
singularities on a trans-Planckian scale. Parenthetically we are encouraged
in our extrapolations by the success of inflationary models, implicitly
referring to these scales\cite{brandenberger}. This reading of B-H entropy
may appear rather natural, as the high curvature region is where new physics
is likely to emerge. However, in passing from the horizon\cite{wald}, where
quantum field theory in curved space-times is expected to work, to the
region close to the classical singularity, in the absence of a full theory
of quantum gravity, we have to rely on heuristic arguments and some guessing
work, which we intend to show can be carried out by rather natural
assumptions. This reading, however, is corroborated by the attractive
features of the Newtonian limit of the model.

Similarly to Ref. \cite{hawkinghertog}, we consider first a simpler fourth
order theory for a scalar field $\phi $, which has the same ghostly behavior
as HD gravity. Its action
\begin{equation}
S=\int d^{4}x\left[ -\phi \square \left( \square -\mu ^{2}\right) \phi
/2-\lambda \phi ^{4}+\alpha \psi ^{\dagger }\psi \phi \right] +S_{mat}\left[
\psi ^{\dagger },\psi \right]
\end{equation}
includes a matter action $S_{mat}$ and an interaction with matter, where $%
\psi ^{\dagger }\psi $ is a shorthand notation for a quadratic scalar
expression in matter fields. Defining
\begin{equation}
\phi _{1}=\left( \square -\mu ^{2}\right) \phi /\mu ,\;\phi _{2}=\square
\phi /\mu ,\;  \label{transformation3}
\end{equation}
the action can be rewritten as
\begin{eqnarray}
&&S\left[ \phi _{1},\phi _{2},\psi ^{\dagger },\psi \right]  \nonumber \\
&=&\int d^{4}x\left[ \frac{1}{2}\phi _{1}\square \phi _{1}-\frac{1}{2}\phi
_{2}\left( \square -\mu ^{2}\right) \phi _{2}-\lambda \left( \frac{\phi
_{2}-\phi _{1}}{\mu }\right) ^{4}+\frac{\alpha }{\mu }\psi ^{\dagger }\psi
\left( \phi _{2}-\phi _{1}\right) \right] +S_{mat}\left[ \psi ^{\dagger
},\psi \right] .  \label{secorder}
\end{eqnarray}
The quadratic term in $\phi _{2}$ has the wrong sign, which classically
means that the energy of this field is negative. Due to the presence of
interactions, energy can flow from negative to positive energy degrees of
freedom, and one can have runaway solutions\cite{hawkinghertog}.

In this model there is a cancellation of all self-energy and vertex
infinities coming from the $\psi ^{\dagger }\psi \phi $ interaction, owing
to the difference in sign between $\phi _{1}$ and $\phi _{2}$ propagators.
This feature, ''analogous to the Pauli-Villars regularization of other field
theories'' \cite{stelle}, is responsible for the improved ultraviolet
behavior in HD gravity \cite{stelle}. A key feature of the non-interacting
theory ($\lambda =\alpha =0$), making it classically viable, can be
considered to be its symmetry under the transformation $\phi
_{2}\longrightarrow -\phi _{2}$, by which symmetrical initial conditions
with $\phi _{2}=\dot{\phi} _{2}=0$ produce symmetrical solutions. If one symmetrizes the
Lagrangian (\ref{secorder}) as it is, in order to extend this symmetry to
the interacting theory, this eliminates the interaction between the ghost
field and the matter altogether and, with it, the mentioned cancellations. A
possible procedure to get a symmetric action while keeping cancellations is
suggested by previous attempts \cite{hawkinghertog} and by the information
loss paradox \cite{unruh2}, both pointing to a non-unitary theory with hidden
degrees of freedom. In particular the most natural way to make the hidden
degrees of freedom ''not ... available as either a net source or a sink of
energy'' \cite{unruh2} is to constraint them to be a copy of observable ones.
Accordingly we introduce a (meta-)matter algebra that is the product of two
copies of the observable matter algebra, respectively generated by the $\psi
^{\dagger },\psi $ and $\tilde{\psi}^{\dagger },\tilde{\psi}$ operators, and
a symmetrized action
\begin{equation}
S_{Sym}=\left\{ S\left[ \phi _{1},\phi _{2},\psi ^{\dagger },\psi \right] +S%
\left[ \phi _{1},-\phi _{2},\tilde{\psi}^{\dagger },\tilde{\psi}\right]
\right\} /2,  \label{symac}
\end{equation}
which is invariant under the symmetry transformation
\begin{equation}
\phi _{1}\longrightarrow \phi _{1},\;\;\phi _{2}\longrightarrow -\phi
_{2},\;\;\psi \longrightarrow \tilde{\psi},\;\;\tilde{\psi}\longrightarrow
\psi .\;
\end{equation}

If the symmetry constraint is imposed on states $\left| \Psi \right\rangle$, i.e. the state space is restricted to those states that are generated from
the vacuum by symmetrical operators, then
\begin{equation}
\left\langle \Psi \right| F\left[ \phi _{2},\psi ^{\dagger },\psi \right]
\left| \Psi \right\rangle =\left\langle \Psi \right| F\left[ -\phi _{2},%
\tilde{\psi}^{\dagger },\tilde{\psi}\right] \left| \Psi \right\rangle
\;\;\forall F.  \label{constraint}
\end{equation}
The allowed states do not give a faithful representation of the original
algebra, which is then larger than the observable algebra. In particular
they cannot distinguish between $F\left[ \psi ^{\dagger },\psi \right] $ and
$F\left[ \tilde{\psi}^{\dagger },\tilde{\psi}\right] $, by which the $\tilde{%
\psi}$ operators are referred to hidden degrees of freedom\cite{unruh2}. On a
classical level $\psi $ and $\tilde{\psi}$ are identified, the $\phi _{2}$
field vanishes and the classical constrained action is that of an ordinary
second order scalar theory interacting with matter:
\begin{equation}
S_{Cl}=\int d^{4}x\left[ \phi _{1}\square \phi _{1}/2-\lambda \left( \phi
_{1}/\mu \right) ^{4}-\alpha \phi _{1}\psi ^{\dagger }\psi /\mu \right]
+S_{mat}\left[ \psi ^{\dagger },\psi \right] .
\end{equation}
%

Consider now the action of a fourth order theory of gravity including matter
\cite{stelle}
\begin{eqnarray}
S &=&S_{G}\left[ g_{\mu \nu }\right] +S_{mat}\left[ g_{\mu \nu },\psi
^{\dagger },\psi \right]  \nonumber \\
&=&-\int d^{4}x\sqrt{-g}\left[ \alpha R_{\mu \nu }R^{\mu \nu }-\beta
R^{2}+R/\left( 16\pi G\right) \right] +\int d^{4}x\sqrt{-g}L_{mat},
\label{hd}
\end{eqnarray}
where $L_{mat}$ denotes the matter Lagrangian density in a generally
covariant form. In terms of the contravariant metric density
\begin{equation}
\sqrt{32\pi G}h^{\mu \nu }=\sqrt{-g}g^{\mu \nu }-\eta ^{\mu \nu },
\end{equation}
the Newtonian limit of the static field is
\begin{equation}
h^{00}\sim 1/r+e^{-\mu _{0}r}/(3r)-4e^{-\mu _{2}r}/(3r),  \label{potential3}
\end{equation}
where $\mu _{0}=[32\pi G(3\beta -\alpha )]^{-1/2}$, $\mu _{2}=[16\pi G\alpha
]^{-1/2}$\cite{stelle}. From Stelle's linearized analysis\cite{stelle}, the
first term in Eq. (\ref{potential3}) corresponds to the graviton, the second
one to a massive scalar and the third one to a negative energy spin-two
field. In fact, in analogy with Eq. (\ref{transformation3}), one can
introduce a transformation from $g_{\mu \nu }$ to a new metric tensor $\bar{g%
}_{\mu \nu }$, a scalar field $\chi $ dilatonically coupled to $\bar{g}_{\mu
\nu }$ and a spin-two field $\phi _{\mu \nu }$, this transformation leading
to the second order form of the action\cite{hindawi}. To be specific,
referring to Ref.\cite{hindawi} (see Eq. (6.9) apart from the matter term),
the action (\ref{hd}) becomes the sum of the Einstein-Hilbert action $S_{EH}$
for $\bar{g}_{\mu \nu }$, an action $S_{gh}$ for $\phi _{\mu \nu }$ and $%
\chi $ coupled to $\bar{g}_{\mu \nu }$, and a matter action $S_{mat}$, with $%
g_{\mu \nu }$ expressed in terms of $\bar{g}_{\mu \nu }$, $\phi _{\mu \nu }$
and $\chi $ (replacing $g_{\mu \nu }$ by $e^{\chi }g_{\mu \nu }$ in Eq.
(4.12) in Ref. \cite{hindawi}):
\begin{eqnarray}
&&S\left[ \bar{g}_{\mu \nu },\phi _{\mu \nu },\chi ,\psi ^{\dagger },\psi %
\right]  \nonumber \\
&=&S_{EH}\left[ \bar{g}_{\mu \nu }\right] +S_{gh}\left[ \bar{g}_{\mu \nu
},\phi _{\mu \nu },\chi \right] +S_{mat}\left[ g_{\mu \nu }(\bar{g}_{\sigma
\tau },\phi _{\sigma \tau },\chi ),\psi ^{\dagger },\psi \right] .
\label{secordergr}
\end{eqnarray}

In $S_{gh}$ the quadratic part in $\phi _{\mu \nu }$ has the wrong sign \cite
{hindawi}. One could symmetrize the action with respect to the
transformation $\phi _{\mu \nu }\rightarrow -\phi _{\mu \nu }$, but this
would eliminate the repulsive term in Eq. (\ref{potential3}), which below
plays a role in avoiding the singularity in gravitational collapse. Like for
the toy model, we double the matter algebra and define the symmetrized
action
\begin{equation}
S_{Sym}=\left\{ S\left[ \bar{g}_{\mu \nu },\phi _{\mu \nu },\chi ,\psi
^{\dagger },\psi \right] +S\left[ \bar{g}_{\mu \nu },-\phi _{\mu \nu },-\chi
,\tilde{\psi}^{\dagger },\tilde{\psi}\right] \right\} /2,  \label{covsymac}
\end{equation}
which is symmetric under the transformation
\begin{equation}
\bar{g}_{\mu \nu }\longrightarrow \bar{g}_{\mu \nu },\;\;\phi _{\sigma \tau
}\rightarrow -\phi _{\sigma \tau },\;\chi \longrightarrow -\chi ,\;\psi
\longrightarrow \tilde{\psi},\;\tilde{\psi}\longrightarrow \psi .\;
\end{equation}
If only symmetric states are allowed, the $\tilde{\psi}$ operators denote
hidden degrees of freedom, as
\begin{equation}
\left\langle \Psi \right| F\left[ \bar{g}_{\mu \nu },\phi _{\mu \nu },\chi
,\psi ^{\dagger },\psi \right] \left| \Psi \right\rangle =\left\langle \Psi
\right| F\left[ \bar{g}_{\mu \nu },-\phi _{\mu \nu },-\chi ,\tilde{\psi}%
^{\dagger },\tilde{\psi}\right] \left| \Psi \right\rangle \;\;\forall F.
\end{equation}
On a classical level $\psi $ and $\tilde{\psi}$ are identified, while the $%
\phi _{\mu \nu }$ and $\chi $ fields vanish and the classical constrained
action is that of \ ordinary matter coupled to ordinary gravity:
\begin{equation}
S_{Cl}\left[ \bar{g}_{\mu \nu },\psi ^{\dagger },\psi \right] =S_{EH}\left[
\bar{g}_{\mu \nu }\right] +S_{mat}\left[ \bar{g}_{\mu \nu },\psi ^{\dagger
},\psi \right] ,  \label{einstein}
\end{equation}
as $S_{gh}\left[ \bar{g}_{\mu \nu },0,0\right] =0$ (Eq. (6.9) in Ref. \cite
{hindawi}) and $g_{\mu \nu }(\bar{g}_{\sigma \tau },0,0)=\bar{g}_{\sigma
\tau }$ (Eq. (4.12) in Ref. \cite{hindawi} with $e^{\chi }g_{\mu \nu }$
replacing $g_{\mu \nu }$).

After the elimination of classical runaway solutions, a further natural step
in assessing the consistency of the theory is the study of its implications
for ordinary laboratory physics. Consider the Newtonian limit with
non-relativistic meta-matter and instantaneous action at a distance. By Eq. (%
\ref{covsymac}), we see that the interactions due to $\bar{g}_{\mu \nu }$
are always attractive, whereas those due to $\phi _{\mu \nu }$ are repulsive
within observable and within hidden meta-matter, as shown by the minus sign
in Eq. (\ref{potential3}), and are otherwise attractive, as the ghostly
character is offset by the difference in sign in its coupling with
observable and hidden meta-matter; the reverse is true for the scalar field $%
\chi $. The corresponding (meta-)Hamiltonian is
\begin{eqnarray}
H_{G} &=&H_{0}[\psi ^{\dagger },\psi ]-\frac{G}{4}\sum_{j,k}m_{j}m_{k}\int
dxdy\frac{:\psi _{j}^{\dagger }(x)\psi _{j}(x)\psi _{k}^{\dagger }(y)\psi
_{k}(y):}{|x-y|}\left( 1+\frac{e^{-\mu _{0}\left| x-y\right| }}{3}-\frac{%
4e^{-\mu _{2}\left| x-y\right| }}{3}\right)   \nonumber \\
&&+H_{0}[\tilde{\psi}^{\dagger },\tilde{\psi}]-\frac{G}{4}%
\sum_{j,k}m_{j}m_{k}\int dxdy\frac{:\tilde{\psi}_{j}^{\dagger }(x)\tilde{\psi%
}_{j}(x)\tilde{\psi}_{k}^{\dagger }(y)\tilde{\psi}_{k}(y):}{|x-y|}\left( 1+%
\frac{e^{-\mu _{0}\left| x-y\right| }}{3}-\frac{4e^{-\mu _{2}\left|
x-y\right| }}{3}\right)   \nonumber \\
&&-\frac{G}{2}\sum_{j,k}m_{j}m_{k}\int dxdy\frac{\psi _{j}^{\dagger }(x)\psi
_{j}(x)\tilde{\psi}_{k}^{\dagger }(y)\tilde{\psi}_{k}(y)}{|x-y|}\left( 1-%
\frac{e^{-\mu _{0}\left| x-y\right| }}{3}+\frac{4e^{-\mu _{2}\left|
x-y\right| }}{3}\right)   \label{metahamiltonian3}
\end{eqnarray}
acting on the product $F_{\psi }\otimes F_{\tilde{\psi}}$ of the Fock spaces
of (the non-relativistic counterparts of) $\psi $ and $\tilde{\psi}$. Two
couples of meta-matter operators $\psi _{j}^{\dagger },\psi _{j}$ and $%
\tilde{\psi}_{j}^{\dagger },\tilde{\psi}_{j}$ \ appear for every particle
species and spin component, while $m_{j}$ is the mass of the $j$-th species
and $H_{0}$ is the gravitationless matter Hamiltonian. The $\tilde{\psi}$
operators obey the same statistics as the corresponding operators $\psi $,
while $[\psi ,\tilde{\psi}]$ $_{-}=[\psi ,\tilde{\psi}^{\dagger }]_{-}=0$.
Tracing out $\tilde{\psi}$ from a symmetrical meta-state evolving according
to the unitary meta-dynamics generated by $H_{G}$ results in a non-Markov
non-unitary physical dynamics for the ordinary matter algebra\cite
{sergio1,sergio2,sergio3}.

Considering, for simplicity, particles of one and the same species, the time
derivative of the matter canonical momentum in a space region $\Omega $ in
the Heisenberg picture reads
\begin{eqnarray}
\frac{d\overrightarrow{p}_{\Omega }}{dt} &=&-i\hslash \frac{d}{dt}%
\int_{\Omega }dx\psi ^{\dagger }(x)\nabla \psi (x)\equiv \left. \frac{d%
\overrightarrow{p}_{\Omega }}{dt}\right| _{G=0}+\vec{F}_{G}=-\frac{i}{%
\hslash }\left[ \overrightarrow{p}_{\Omega },H_{0}[\psi ^{\dagger },\psi ]%
\right]  \nonumber \\
&&+\frac{G}{2}m^{2}\int_{\Omega }dx\psi ^{\dagger }(x)\psi (x)\nabla
_{x}\int_{R^{3}}dy\frac{\tilde{\psi}^{\dagger }(y)\tilde{\psi}(y)}{\left|
x-y\right| }\left( 1-\frac{e^{-\mu _{0}\left| x-y\right| }}{3}+\frac{%
4e^{-\mu _{2}\left| x-y\right| }}{3}\right)  \nonumber \\
&&+\frac{G}{2}m^{2}:\int_{\Omega }dx\psi ^{\dagger }(x)\psi (x)\nabla
_{x}\int_{R^{3}}dy\frac{\psi ^{\dagger }(y)\psi (y)}{\left| x-y\right| }%
\left( 1+\frac{e^{-\mu _{0}\left| x-y\right| }}{3}-\frac{4e^{-\mu _{2}\left|
x-y\right| }}{3}\right) :.
\end{eqnarray}
The expectation of the gravitational force can be written as
\begin{eqnarray}
\left\langle \vec{F}_{G}\right\rangle &=&\frac{G}{2}m^{2}\left\langle
\int_{\Omega }dx\psi ^{\dagger }(x)\psi (x)\nabla _{x}\int_{\Omega }dy\frac{%
\tilde{\psi}^{\dagger }(y)\tilde{\psi}(y)}{\left| x-y\right| }\left( 1-\frac{%
e^{-\mu _{0}\left| x-y\right| }}{3}+\frac{4e^{-\mu _{2}\left| x-y\right| }}{3%
}\right) \right\rangle  \nonumber \\
&&+\frac{G}{2}m^{2}\left\langle \int_{\Omega }dx\psi ^{\dagger }(x)\psi
(x)\nabla _{x}\int_{R^{3}\backslash \Omega }dy\frac{\tilde{\psi}^{\dagger
}(y)\tilde{\psi}(y)}{\left| x-y\right| }\left( 1-\frac{e^{-\mu _{0}\left|
x-y\right| }}{3}+\frac{4e^{-\mu _{2}\left| x-y\right| }}{3}\right)
\right\rangle  \nonumber \\
&&+\frac{G}{2}m^{2}\left\langle :\int_{\Omega }dx\psi ^{\dagger }(x)\psi
(x)\nabla _{x}\int_{\Omega }dy\frac{\psi ^{\dagger }(y)\psi (y)}{\left|
x-y\right| }\left( 1+\frac{e^{-\mu _{0}\left| x-y\right| }}{3}-\frac{%
4e^{-\mu _{2}\left| x-y\right| }}{3}\right) :\right\rangle  \nonumber \\
&&+\frac{G}{2}m^{2}\left\langle \int_{\Omega }dx\psi ^{\dagger }(x)\psi
(x)\nabla _{x}\int_{R^{3}\backslash \Omega }dy\frac{\psi ^{\dagger }(y)\psi
(y)}{\left| x-y\right| }\left( 1+\frac{e^{-\mu _{0}\left| x-y\right| }}{3}-%
\frac{4e^{-\mu _{2}\left| x-y\right| }}{3}\right) \right\rangle ,
\end{eqnarray}
where, on allowed states, the first term vanishes for the antisymmetry of
the kernel

$\nabla _{x}\left[ \left( 1-e^{-\mu _{0}|x-y|}/3+4e^{-\mu
_{2}|x-y|}/3\right) /\left| x-y\right| \right] $ and the symmetry of the
state, while the third one vanishes just as a consequence of the
antisymmetry of the corresponding kernel. We can approximate $\left\langle
\psi ^{\dagger }(x)\psi (x)\tilde{\psi}^{\dagger }(y)\tilde{\psi}%
(y)\right\rangle $ and $\left\langle \psi ^{\dagger }(x)\psi (x)\psi
^{\dagger }(y)\psi (y)\right\rangle $ respectively with $\left\langle \psi
^{\dagger }(x)\psi (x)\right\rangle \left\langle \tilde{\psi}^{\dagger }(y)%
\tilde{\psi}(y)\right\rangle $ and $\left\langle \psi ^{\dagger }(x)\psi
(x)\right\rangle \left\langle \psi ^{\dagger }(y)\psi (y)\right\rangle $ ,
as $x\in \Omega $ and $y\in R^{3}\backslash \Omega $. Finally, as $%
\left\langle \tilde{\psi}^{\dagger }(y)\tilde{\psi}(y)\right\rangle
=\left\langle \psi ^{\dagger }(y)\psi (y)\right\rangle $, we get
\begin{equation}
\left\langle \vec{F}_{G}\right\rangle \simeq Gm^{2}\int_{\Omega
}dx\left\langle \psi ^{\dagger }(x)\psi (x)\right\rangle \nabla
_{x}\int_{R^{3}\backslash \Omega }dy\left\langle \psi ^{\dagger }(y)\psi
(y)\right\rangle /\left| x-y\right| ,
\end{equation}
as for the traditional Newton interaction between observable degrees of
freedom only, consistently with the classical equivalence of the original
theory to Einstein gravity.

As to the quantum aspects of the present Newtonian model, a closely related
model was analyzed in Ref.s \cite{sergio1,sergio2,sergio3}. Actually, if in Ref. \cite
{sergio1,sergio2,sergio3} $H[\psi ^{\dagger },\psi ]$ is meant to be the sum of $%
H_{0}[\psi ^{\dagger },\psi ]$ and the normal ordered interaction within
observable matter in Eq. (\ref{metahamiltonian3}) above, and analogously for
the hidden meta-matter, the two models differ only for the kernel in the
interaction between observable and hidden meta-matter. The main results,
which stay qualitatively unchanged, are the following. For the center of
mass wave function of a homogeneous body of mass $M$ and linear dimensions $%
R $, effective gravitational self-interactions lead to a localization length
$\Lambda \sim (\hslash ^{2}R^{3}/GM^{3})^{1/4}$, as soon as it is small with
respect to $R$. This produces a rather sharp threshold that, for ordinary
densities $\sim 10^{24}m_{p}/cm^{3}$, where $m_{p}$ denotes the proton mass,
is around $10^{11}m_{p}$, below which the effects of the effective
gravitational self-interactions are irrelevant\cite{sergio1,sergio2,sergio3}. A localized
state slowly evolves, with times $\sim 10^{3}\sec $, for ordinary densities,
into a delocalized ensemble of localized states\cite{sergio1,sergio2,sergio3}, this
entropic spreading replacing the wave function spreading of ordinary QM and
the unphysical stationary localized states of the nonlinear unitary
Schroedinger-Newton (S-N) model \cite{christian,penrose1,moroz,kumar}. As to
an unlocalized pure state of a body above threshold, it rapidly gets
localized, within times $\sim 10^{20}(M/m_{p})^{-5/3}\sec $, under
reasonable assumptions on the initial state, into such an ensemble\cite
{sergio1,sergio2,sergio3}. This gives a well-defined dynamical model for gravity-induced
decoherence, to be compared with purely numerological estimates\cite
{karolyhazy1,hawking3,diosi,ghirardi2,penrose,squires1,ellis2,anandan} and
which allows us to address physically relevant problems, like the
characterization of gravitationally decoherence free states of the physical
operator algebra\cite{defilippo0}. It is worthwhile to remark that, in spite
of the presence of the masses $\mu _{i},i=0,2$ (actually $\hslash c\mu _{i}$%
), the Newtonian limit has, for all practical purposes, no free parameter,
as to ordinary laboratory physics, if as usual they are assumed to be of the
order of the Planck mass, which is equivalent to take the limit $\mu
_{i}\rightarrow \infty $.

If the traditional Hamiltonian includes the Newton interaction, there are
extremely small violations of energy conservation, as only the
meta-Hamiltonian $H_{G}$ is strictly conserved. These fluctuations are
consistent with the assumption that an eigenstate of \ the traditional
Hamiltonian may evolve towards a microcanonical mixed state with an energy
dispersion around the original energy, which, though irrelevant on a
macroscopic scale, paves the way for the possibility that the thermodynamic
entropy of a closed system may be identified with its von Neumann entropy
\cite{kay}. This is not irrelevant, if ''...in order to gain a better
understanding of the degrees of freedom responsible for black hole entropy,
it will be necessary to achieve a deeper understanding of the notion of
entropy itself. Even in flat space-time, there is far from universal
agreement as to the meaning of entropy -- particularly in quantum theory --
and as to the nature of the second law of thermodynamics''\cite{wald}. Of
course the reversibility of the unitary meta-dynamics makes entropy decrease
conceivable too\cite{schulman}, so that a derivation of the entropy-growth
for a closed system in the present context must have recourse to the choice
of suitable initial conditions, like unentanglement between the observable
and the hidden algebras\cite{kay}. While the assumption of special initial
conditions dates back to Boltzmann, only a non-unitary dynamics makes it a
viable starting point, within a quantum context, for the microscopic
derivation of the second law of thermodynamics, in terms of von Neumann
entropy, without renouncing strict isolation\cite{gemmer}.

Emboldened by the mentioned bonuses coming from the Newtonian limit of our
model, we now try to apply it to gravitational collapse. First evaluate
within our model the linear dimension of a collapsed matter lump, replacing
the classical singularity. In order to do that we boldly use Eq. (\ref
{metahamiltonian3}) for lengths smaller than $\mu _{0}^{-1}$ and $\mu
_{2}^{-2}$, namely in the limit $\mu _{0},\mu _{2}\rightarrow 0$. This
corresponds to the replacement of our meta-Hamiltonian with the
meta-Hamiltonian in Ref. \cite{sergio1,sergio2,sergio3}, where there is no gravitational
interaction within observable and within hidden matter, while there is a
Newton interaction between observable and hidden matter. This interaction is
effective in lowering the gravitational energy of a matter lump as far as
the localization length $\Lambda =(\hslash ^{2}R^{3}/GM^{3})^{1/4}$ is
fairly smaller than the lump radius $R$\cite{sergio1,sergio2,sergio3}. The highest
possible density then corresponds roughly to $\Lambda =R$, namely to
\begin{equation}
R=\hslash ^{2}/\left( GM^{3}\right) .  \label{minimal}
\end{equation}
In fact, below the localization threshold, only the interactions within
observable (and hidden meta-) matter are effective in collapsing matter,
but, in the considered limit $\mu _{0},\mu _{2}\rightarrow 0$, they vanish.
This parenthetically shows that the following discussion depends crucially,
not only on the doubling of the matter degrees of freedom, but also on the
inclusion of the repulsive interactions of HD gravity.

As to the space-time geometry, the Schwarzschild metric in ingoing
Eddington-Finkelstein coordinates ($v,r,\theta ,\phi $) covers the two
regions of the Kruskal maximal extension that are relevant to gravitational
collapses\cite{townsend}:
\begin{equation}
ds^{2}=-\left[ 1-2MG/\left( rc^{2}\right) \right] dv^{2}+2drdv+r^{2}\left[
d\theta ^{2}+\sin ^{2}\theta d\phi ^{2}\right] .
\end{equation}
If, in the region beyond the horizon we put $x=v-\int dr\left[ 1-2MG/\left(
rc^{2}\right) \right] ^{-1}$, then
\begin{equation}
ds^{2}=\left[ 1-2MG/\left( rc^{2}\right) \right] ^{-1}dr^{2}-\left[
1-2MG/\left( rc^{2}\right) \right] dx^{2}+r^{2}\left[ d\theta ^{2}+\sin
^{2}\theta d\phi ^{2}\right] ,  \label{xrmetric}
\end{equation}
where we see that beyond the horizon $r$ may be regarded as a time variable
\cite{townsend}.

If we trust (\ref{minimal}) as the minimal length involved in the collapse,
we are led to assume that a full theory of quantum gravity should include a
mechanism regularizing the singularity at $r=0$ by means of that minimal
length. In particular, to characterize the region occupied by the collapsed
lump, consider that for time-like geodesics at constant $\theta $ and $\phi $
one can show that $\left| dx/dr\right| \sim r^{3/2}$ \ as $r\rightarrow 0$.
This implies that, the $x$ coordinate difference $\Delta x$\ of two material
points has a well defined limit as $r\rightarrow 0$, by which it is natural
to assume that the $x$ width of the collapsed matter lump is $\Delta x\sim R$%
. As to the apparent inconsistency of matter occupying just a finite $\Delta
x$ interval with $\partial /\partial x$ being a Killing vector, one should
expect on sub-Planckian scales substantial quantum corrections to the
Einstein equations that the model gives on a classical level (\ref{einstein}%
), with the dilaton and the ghost fields, though vanishing in the average,
playing a crucial role. On the other hand we are proceeding according to the
usual assumption, or fiction, of QM on the existence of a global time
variable, at least in the region swept by the lump. In fact the most natural
way to regularize (\ref{xrmetric}) is to consider it as an approximation for
$r>R$ of a regular metric, whose coefficients for $r\rightarrow 0$
correspond to the ones in (\ref{xrmetric}) with $r=R$, in which case there
is no obstruction in extending the metric to $r<0$, where taking constant
coefficients makes $\partial /\partial r$ a time-like Killing vector. As a
consequence, the relevant space metric in the region swept by the collapsed
lump is
\begin{equation}
ds_{SPACE}^{2}\sim 2MG/\left( Rc^{2}\right) dx^{2}+R^{2}\left[ d\theta
^{2}+\sin ^{2}\theta d\phi ^{2}\right] .
\end{equation}
The volume of the collapsed matter lump is then:
\begin{equation}
V\sim R^{2}\Delta x\sqrt{MG/\left( Rc^{2}\right) }=\left[ \hslash
^{2}/\left( GM^{3}\right) \right] ^{5/2}\sqrt{MG/c^{2}}=\hslash
^{5}M^{-7}/\left( G^{2}c\right) .
\end{equation}
According to the above view, thermodynamical equilibrium is reached, due to
the gravitational interaction generating entanglement between the observable
and hidden meta-matter, by which the matter state is a microcanonical
ensemble corresponding to the energy
\begin{equation}
E=Mc^{2}+GM^{2}/R=Mc^{2}+GM^{2}\left[ GM^{3}/\hslash ^{2}\right] \sim
G^{2}M^{5}/\hslash ^{2},\ \ if\ \ M\gg M_{P},
\end{equation}
where $M_{P}=\sqrt{\hslash c/G}$ is the Planck mass, and to the energy
density
\begin{equation}
\varepsilon =E/V\sim G^{4}cM^{12}/\hslash ^{7}.
\end{equation}
We first treat the collapsed lump as a three-dimensional bulk in spite of
the huge dilation factor in the $x$ direction. Since this energy density
corresponds to a very high temperature, not to be mistaken for the Hawking
temperature, the matter can be represented by massless fields, whose
equilibrium entropy is given by
\begin{equation}
S\sim \left( K_{B}/\not{h}^{3/4}c^{3/4}\right) \varepsilon
^{3/4}V=GM^{2}K_{B}/\left( \hslash c\right), \label{entropy4}
\end{equation}
that of course coincides, apart for $K_{B}$ set suitably to $1$ and numerical factor of $O(1)$, with the usual formula $\frac{1}{4}\frac{A}{l_p^2}$ where $A$ is the horizon area and  $l_p$ the Planck length. Of course this result can be trusted at most for its order of magnitude, the
uncertainty in the number of species being just one part of an unknown
numerical factor. With this proviso, common to other approaches\cite{wald},
Eq. (\ref{entropy4}) agrees with B-H entropy. As this result contains four
dimensional constants, it is more than pure numerology: from a dimensional
viewpoint one could replace the dimensionless quantity $GM^{2}/\hslash c$ in
Eq. (\ref{entropy4}) by any arbitrary function of $GM^{2}/\hslash c$. For
instance, if we ignored that the space geometry near the smoothed
singularity is not euclidean and then we assumed that $V$ $\sim R^{3}$, we
would get the entropy to be independent from the black hole mass, since the
increase of the entropy density with the mass would be offset by the
shrinking of the volume.

One could object, against the above derivation, that the collapsed matter
lump, for the presence of the dilation factor along the $x$ direction, is
more like a one-dimensional string-like structure of transverse dimensions $%
\sim $\ $R$ along $\theta $ and $\phi $ and length $L=(MG/Rc^{2})^{1/2}R\gg
R $. If we treat it like this, for the linear density of entropy we have $%
s\varpropto \varepsilon ^{1/2}$ and for the length $L\varpropto M^{-1}$, by
which $S=Ls\varpropto L(E/L)^{1/2}\varpropto M^{2}$, which agrees with the
previous result and with B-H entropy.

Finally, if we give for granted that a future theory of quantum gravity will
account for black hole evaporation, we can connect the temperature
\begin{equation}
T\sim \sqrt[4]{\varepsilon h^{3}c^{3}}/K_{B}\sim cGM^{3}/K_{B}\hslash
\end{equation}
of our collapsed matter lump with the temperature of the radiation at
infinity. If we model radiation by massless fields, emitted for simplicity
at a constant temperature as we are interested just in orders of magnitude,
this temperature is defined in terms of he ratio $E_{\infty }/S_{\infty }$
of its energy $E_{\infty }$ and its entropy $S_{\infty }$. It is natural to
assume that, ''once'' thermodynamical equilibrium is reached due to the
highly non-unitary dynamics close to the classical singularity, no entropy
production occurs during evaporation, by which $S_{\infty }=S$. Then, if $%
E_{\infty }=Mc^{2}$ is the energy of the total Hawking radiation spread over
a very large space volume, its temperature agrees with Hawking temperature,
i.e. \
\begin{equation}
T_{\infty }=\left( E_{\infty }/E\right) T\sim \left( c^{3}\hslash
/MGK_{B}\right).
\end{equation}

\section{Quantum measurement and rigorous definition of coarse graining entropy}
\label{section7}

In this Section it is shown that the notion of coarse graining entropy, often
taken as the starting point in dealing with the quantum foundations of the
second law of thermodynamics \cite{vNeumann}, can be easily connected with
the present approach. Consider, for simplicity, a non-degenerate physical
state
\begin{equation}
\rho _{Ph}=\sum_{j}p_{j}\left| j\right\rangle \left\langle j\right|
,\;p_{j}\in R,\;\;p_{j}=p_{k}\Rightarrow j=k.\;
\end{equation}
The most general pure meta-state vector giving rise to $\rho _{ph}$ is
\begin{equation}
\left| \left| \Psi \right\rangle \right\rangle _{{\bf \varphi }%
}=\sum_{j}e^{i\varphi _{j}}\sqrt{p_{j}}\left| j\right\rangle \left|
j\right\rangle ,  \label{microstate5}
\end{equation}
where $\left| j\right\rangle \left| j\right\rangle $ denotes the tensor
product of two corresponding vectors in the two Fock spaces and the $\varphi
_{j}\in \lbrack 0,2\pi \lbrack $ are arbitrary real parameters. The
indistinguishability of the corresponding meta-states, due to the
restriction of the physical algebra, induces in the meta-state space an
unambiguous coarse graining, at variance with the rather vague one in the
traditional approaches. To be specific, it is natural to introduce the
macro-meta-state
\begin{equation}
\rho _{CG}\equiv \int \prod_{j}\frac{d\varphi _{j}}{2\pi }\left| \left| \Psi
\right\rangle \right\rangle _{{\bf \varphi }}\left\langle \left\langle \Psi
\right| \right| =\sum_{j}p_{j}\left| j\right\rangle \left| j\right\rangle
\left\langle j\right| \left\langle j\right|,
\end{equation}
corresponding to the equiprobability of the micro-meta-states $\left| \left|
\Psi \right\rangle \right\rangle _{{\bf \varphi }}\left\langle \left\langle
\Psi \right| \right| .$ The corresponding coarse graining entropy is
\begin{equation}
S_{CG}=-K_{B}Tr\left[ \rho _{CG}\ln \rho _{CG}\right] =-K_{B}\sum_{j}p_{j}%
\ln p_{j},
\end{equation}
which coincides with the von Neumann entropy of the physical state $\rho
_{Ph}$.

Vice versa, if we assume that a specific pure meta-state $\left| \left| \Psi
\right\rangle \right\rangle$ is given, the Schmidt decomposition theorem
allows us to write it in terms of orthonormal vectors as
\begin{equation}
\left| \left| \Psi \right\rangle \right\rangle =\sum_{j}\sqrt{p_{j}}\left|
j\right\rangle \left| j^{\prime }\right\rangle ,
\end{equation}
with the $p_{j}$ positive, for simplicity distinct, real numbers. By the
symmetry constraint on the meta-state space one can choose the relative
phases in such a way that $\left| j\right\rangle $ and $\left| j^{\prime
}\right\rangle $ can be taken as corresponding vectors in the two Fock
spaces, thus reproducing $\left| \left| \Psi \right\rangle \right\rangle _{{\bf \varphi }}$ in eq. (\ref{microstate5}) for ${\bf \varphi }=0$. Although this amounts to the knowledge of a definite microstate, the entropy of the
corresponding physical state $\rho _{Ph}$ is non-vanishing and coincides
with the coarse graining entropy of the corresponding macrostate $\rho _{CG}$. This shows the objective and non-conventional character of the notion of entropy in the present approach, since it does not depend on a subjective
characterization based on the notion of a macroscopic observer \cite{vNeumann}.

In an interaction representation of the ordinary Newtonian limit, where the
free meta-Hamiltonian is $H[\psi ^{\dagger },\psi ]+H[\tilde{\psi}^{\dagger},\tilde{\psi}]$, the time evolution of an initially untangled meta-state $\left| \left| \tilde{\Phi}(0)\right\rangle \right\rangle $\ is represented
by
\begin{eqnarray}
\left| \left| \tilde{\Phi}(t)\right\rangle \right\rangle &=&{\it T}\exp %
\left[ \frac{i}{\hslash }Gm^{2}\int dt\int dxdy\frac{\psi ^{\dagger
}(x,t)\psi (x,t)\tilde{\psi}^{\dagger }(y,t)\tilde{\psi}(y,t)}{|x-y|}\right]
\left| \left| \tilde{\Phi}(0)\right\rangle \right\rangle  \nonumber \\
&\equiv &U(t)\left| \left| \tilde{\Phi}(0)\right\rangle \right\rangle \equiv
U(t)\left| \Phi (0)\right\rangle _{\psi }\otimes \left| \Phi
(0)\right\rangle _{\chi }.  \label{evolvedmetastate5}
\end{eqnarray}
Then, by a Stratonovich-Hubbard transformation\cite{negele}, we can rewrite $U(t)$ as
\begin{eqnarray}
U(t) &=&\int {\it D}\left[ \varphi _{1},\varphi _{2}\right] \exp \frac{ic^{2}%
}{2\hslash }\int dtdx\left[ \varphi _{1}\nabla ^{2}\varphi _{1}-\varphi
_{2}\nabla ^{2}\varphi _{2}\right]  \nonumber \\
&&{\it T}\exp \left[ -i\frac{mc}{\hslash }\sqrt{2\pi G}\int dtdx\left[
\varphi _{1}(x,t)+\varphi _{2}(x,t)\right] \psi ^{\dagger }(x,t)\psi (x,t)%
\right]  \nonumber \\
&&{\it T}\exp \left[ -i\frac{mc}{\hslash }\sqrt{2\pi G}\int dtdx\left[
\varphi _{1}(x,t)-\varphi _{2}(x,t)\right] \tilde{\psi}^{\dagger }(x,t)%
\tilde{\psi}(x,t)\right]  \label{stratonovich5}
\end{eqnarray}
namely as a functional integral over two auxiliary real scalar fields $%
\varphi _{1}$ and $\varphi _{2}$.

The physical state corresponding to the meta-state (\ref{evolvedmetastate5})
is given by
\begin{equation}
\rho _{Ph}(t)\equiv Tr_{\tilde{\psi}}\left| \left| \tilde{\Phi}%
(t)\right\rangle \right\rangle \left\langle \left\langle \tilde{\Phi}%
(t)\right| \right| =\sum_{k}\;\;\;_{\tilde{\psi}}\left\langle k\right|
\left| \left| \tilde{\Phi}(t)\right\rangle \right\rangle \left\langle
\left\langle \tilde{\Phi}(t)\right| \right| \left| k\right\rangle _{\tilde{%
\psi}}
\end{equation}
and, by using Eq. (\ref{stratonovich5}), we can write
\[
_{\tilde{\psi}}\left\langle k\right| \left| \left| \tilde{\Phi}%
(t)\right\rangle \right\rangle =\int {\it D}\left[ \varphi _{1},\varphi _{2}%
\right] \exp \frac{ic^{2}}{2\hslash }\int dtdx\left[ \varphi _{1}\nabla
^{2}\varphi _{1}-\varphi _{2}\nabla ^{2}\varphi _{2}\right]
\]
\begin{eqnarray}
&&_{\tilde{\psi}}\left\langle k\right| {\it T}\exp \left[ -i\frac{mc}{%
\hslash }\sqrt{2\pi G}\int dtdx\left[ \varphi _{1}(x,t)-\varphi _{2}(x,t)%
\right] \tilde{\psi}^{\dagger }(x,t)\tilde{\psi}(x,t)\right] \left| \Phi
(0)\right\rangle _{\tilde{\psi}}  \nonumber \\
&&{\it T}\exp \left[ -i\frac{mc}{\hslash }\sqrt{2\pi G}\int dtdx\left[
\varphi _{1}(x,t)+\varphi _{2}(x,t)\right] \psi ^{\dagger }(x,t)\psi (x,t)%
\right] \left| \Phi (0)\right\rangle _{\tilde{\psi}}.
\end{eqnarray}
Then the final expression for the physical state at time $t$ is given by
\[
\rho _{Ph}(t)=\int {\it D}\left[ \varphi _{1},\varphi _{2},\varphi
_{1}^{\prime },\varphi _{2}^{\prime }\right] \exp \frac{ic^{2}}{2\hslash }%
\int dtdx\left[ \varphi _{1}\nabla ^{2}\varphi _{1}-\varphi _{2}\nabla
^{2}\varphi _{2}-\varphi _{1}^{\prime }\nabla ^{2}\varphi _{1}^{\prime
}+\varphi _{2}^{\prime }\nabla ^{2}\varphi _{2}^{\prime }\right]
\]
\begin{eqnarray}
&&_{\psi }\left\langle \Phi (0)\right| {\it T}^{-1}\exp \left[ i\frac{mc}{%
\hslash }\sqrt{2\pi G}\int dtdx\left[ \varphi _{1}^{\prime }-\varphi
_{2}^{\prime }\right] \psi ^{\dagger }\psi \right]  \nonumber \\
&&{\it T}\exp \left[ -i\frac{mc}{\hslash }\sqrt{2\pi G}\int dtdx\left[
\varphi _{1}-\varphi _{2}\right] \psi ^{\dagger }\psi \right] \left| \Phi
(0)\right\rangle _{\psi }  \nonumber \\
&&{\it T}\exp \left[ -i\frac{mc}{\hslash }\sqrt{2\pi G}\int dtdx\left[
\varphi _{1}+\varphi _{2}\right] \psi ^{\dagger }\psi \right] \left| \Phi
(0)\right\rangle _{\psi }  \nonumber \\
&&_{\psi }\left\langle \Phi (0)\right| {\it T}^{-1}\exp \left[ i\frac{mc}{%
\hslash }\sqrt{2\pi G}\int dtdx\left[ \varphi _{1}^{\prime }+\varphi
_{2}^{\prime }\right] \psi ^{\dagger }\psi \right]  \label{alternative5}
\end{eqnarray}
where, due to the constraint on the meta-state space, $\tilde{\psi}$
operators were replaced by $\psi $ operators, and the meta-state vector $%
\left| \Phi (0)\right\rangle _{\tilde{\psi}}$ by $\left| \Phi
(0)\right\rangle _{\psi }$. This expression can even be taken as an
independent equivalent definition of the non-unitary dynamics, free from any
reference to the extended algebra including unobservable degrees of freedom.

Consider an initial linear, for simplicity orthogonal, superposition of $N$
localized states of a macroscopic body, existing, as shown above, as pure
states corresponding to unentangled bound meta-states for bodies of ordinary
density and a mass $M$ higher than $\sim 10^{11}m_{p}$\cite{sergio3}:
\begin{equation}
\left| \Phi (0)\right\rangle =\frac{1}{\sqrt{N}}\sum_{j=1}^{N}\left|
z_{j}\right\rangle  \label{superposition5}
\end{equation}
where $\left| z\right\rangle $ represents a localized state centered in $z$.
We consider the localized states as approximate eigenstates of the particle
density operator, i.e. $\psi ^{\dagger }(x,t)\psi (x,t)\left| z\right\rangle
\simeq n(x-z)\left| z\right\rangle $, where time dependence is irrelevant,
consistently with these states being stationary both in the gravity-free and
in the interacting Schr\"{o}dinger pictures apart from a slow spreading,
which, as shown below, is much slower than the computed time for wave
function reduction.

According to Eq. (\ref{alternative5}), the density matrix elements are then
given by
\begin{eqnarray}
&&\left\langle z_{h}\right| \rho _{Ph}(t)\left| z_{k}\right\rangle  \nonumber
\\
&=&\int {\it D}\left[ \varphi _{1},\varphi _{2},\varphi _{1}^{\prime
},\varphi _{2}^{\prime }\right] \exp \frac{ic^{2}}{2\hslash }\int dtdx\left[
\varphi _{1}\nabla ^{2}\varphi _{1}-\varphi _{2}\nabla ^{2}\varphi
_{2}-\varphi _{1}^{\prime }\nabla ^{2}\varphi _{1}^{\prime }+\varphi
_{2}^{\prime }\nabla ^{2}\varphi _{2}^{\prime }\right]  \nonumber \\
&&\frac{1}{N^{2}}\sum_{j=1}^{N}\exp \left[ -i\frac{mc}{\hslash }\sqrt{2\pi G}%
\int dtdx\left[ \left[ \varphi _{1}-\varphi _{2}\right] n(x-z_{j})-\left[
\varphi _{1}^{\prime }-\varphi _{2}^{\prime }\right] n(x-z_{j})\right] %
\right]  \nonumber \\
&&\exp \left[ -i\frac{mc}{\hslash }\sqrt{2\pi G}\int dtdx\left[ \left[
\varphi _{1}+\varphi _{2}\right] n(x-z_{h})-\left[ \varphi _{1}^{\prime
}+\varphi _{2}^{\prime }\right] n(x-z_{k})\right] \right]
\end{eqnarray}
and, after integrating out the scalar fields,
\begin{equation}
\left\langle z_{h}\right| \rho _{Ph}(t)\left| z_{k}\right\rangle =\frac{1}{%
N^{2}}\sum_{j=1}^{N}\exp \frac{i}{\hslash }Gm^{2}t\int dxdy\left[ \frac{%
n(x-z_{j})n(y-z_{h})}{|x-y|}-\frac{n(x-z_{j})n(y-z_{k})}{|x-y|}\right]
\label{coherences5}
\end{equation}
which shows that, while diagonal elements are given by $\left\langle
z_{h}\right| \rho _{Ph}(t)\left| z_{h}\right\rangle =1/N$, the coherences,
under reasonable assumptions on the linear superposition in Eq. (\ref
{superposition5}) of a large number of localized states, approximately
vanish, due to the random phases in Eq. (\ref{coherences5}). This makes the
state $\rho _{Ph}(t)$, for times $t\gtrsim T_{G}\sim
10^{20}(M/m_{p})^{-5/3}\sec $, which are consistently short with respect to
the time of the entropic spreading $\sim 10^{3}\sec $, equivalent to an
ensemble of localized states:
\begin{equation}
\rho _{Ph}(t)\simeq \frac{1}{N}\sum_{j=1}^{N}\left| z_{j}\right\rangle
\left\langle z_{j}\right| .
\end{equation}
It is worthwhile to remark that the extremely short localization time of a
macroscopic body may make its unlocalized states unobservable for all
practical purposes. The above analysis is also supported by numerical
evidence independently from the particular assumptions made here on the
initial unlocalized state (\ref{superposition5}) \cite{sergioFil2}. In such a way one gets a gravity-induced
dynamical reduction of the wave function, which up to now was assumed to
follow, possibly, from a future theory of quantum gravity \cite{karolyhazy1}.
It is worthwhile to remark that the order of magnitude of decoherence times
in Eq. (\ref{coherences5}) agrees with the one obtained by previous
numerological arguments for gravity-induced localization \cite{penrose}: ``Although a detailed estimate of $T_{G}$ would require a full theory of quantum gravity... it is reasonable to expect that for non-relativistic
systems ..." \cite{melko}. \ \ What is new here in this regard is a fully
defined dynamical model without any free parameter, which in principle
allows for the explicit evaluation of any physically relevant quantity and
for addressing crucial questions like the search for
(gravitational-)decoherence free states of the physical operator algebra \cite{defilippo0}.

To be more specific, we have derived the first unified model for Newtonian
gravity and gravity-induced decoherence. If the states $\left|
z_{j}\right\rangle $ in Eq. (\ref{superposition5}) are the pointer states of
a measurement apparatus and $\left| e_{j}\right\rangle $ are the measurement
eigenstates of a microscopic system, the product state
\begin{equation}
\left| z_{0}\right\rangle \otimes \sum_{j}c_{j}\left| e_{j}\right\rangle
\end{equation}
according to the traditional von Neumann model for the interaction between
the two systems, is transformed into an entangled state \cite{vNeumann}
\begin{equation}
\sum_{j}c_{j}\left| z_{j}\right\rangle \otimes \left| e_{j}\right\rangle.
\end{equation}

Obviously our previous analysis of the effect of the gravitational
(self-)interaction on the quantum motion of the macroscopic body is not
affected by the presence of the microscopic system, by which the reduction
of the wave function occurs:
\begin{equation}
\sum_{j,k}c_{j}\bar{c}_{k}\left| z_{j}\right\rangle \otimes \left|
e_{j}\right\rangle \left\langle e_{k}\right| \otimes \left\langle
z_{k}\right| \longrightarrow \sum_{j}\left| c_{j}\right| ^{2}\left|
z_{j}\right\rangle \otimes \left| e_{j}\right\rangle \left\langle
e_{j}\right| \otimes \left\langle z_{j}\right| .  \label{reduction5}
\end{equation}
Of course one can look in principle for a collapse model \cite{ghirardi2,pearle2} in terms of a stochastic dynamics for pure states, which, when averaged, leads to Eq. (\ref{reduction5}). Apart, in principle, from the
non uniqueness of the stochastic realization \cite{pearle2}, stochastic models
can certainly be useful as computational tools \cite{carmichael}. However the
view advocated here considers density matrices arising from gravitational
decoherence as the fundamental characterization of the system state and not
just as a bookkeeping tool for statistical uncertainties. The fact that the
apparent uniqueness of the measurement result seems to imply a real collapse
is perhaps more an ontological than a physical problem, and presumably, if
one likes it, that can be addressed by a variant of the many-world
interpretation \cite{everett,gellmann}.

\section{Conclusions}
\label{section8}

In spite to what has been already achieved, much remains to be done.

To verify more convincingly thermalization one needs to examine more general and
realistic systems, beginning from an anharmonic crystal where it is possible to get
thermalization even starting from a state with a single phonon due to process of a
single phonon decaying in two phonons.

It should be found a more rigorous treatment of black hole singularities regularization.

Finally a detailed analysis of perturbative expansion of non-unitary HD Gravity
should be addressed in order to prove finiteness of self-energy and vertex functions.

\section*{Acknowledgements}

SDF thanks Mario Salerno for finding the best at the time hardsoft eye tracker
system allowing him, for instance, to give the guidelines for this paper and the
Dipartimento di Fisica di Salerno for financing its acquisition; he thanks Joseph
Quartieri with his engineering student F S Giordano for fixing severe hardsoft
problems.

Finally SDF wants to express his gratitude to all his collaborators and in
particular AN who allowed him finally to present motivations, assumptions and
conceptual framework behind his non-unitary gravity that he could not present in his
first papers written in a hurry for his awareness that in a little time he could not
use pc keyboard or write on paper.

\begin{appendices}
\numberwithin{equation}{section}

\section{Calculation of $C(G,V)$}
\label{coeffCalc}

Let's calculate the gravitational interaction between two phonons, supposed
to fill homogeneously the crystal of cubic form. Their mass are expressed by
\begin{equation}
m_{i}=\frac{\hbar \omega _{\mathbf{k}_{i}}}{c^{2}},\ \ \ \ \ \ \ \ \ \
i=1,2\
\end{equation}
while their gravitational interaction energy is
\begin{equation}
U_{G}=-m_{1}m_{2}C(G,V)=-m_{1}m_{2}\ G\ V^{-1/3}\ \Im ,
\end{equation}
where
\begin{equation}
\Im =\int\limits_{0}^{1}d\xi \int\limits_{0}^{1}d\eta
\int\limits_{0}^{1}d\zeta \int\limits_{0}^{1}d\xi ^{\prime
}\int\limits_{0}^{1}d\eta ^{\prime }\int\limits_{0}^{1}d\zeta ^{\prime }%
\frac{1}{\sqrt{\left( \xi -\xi ^{\prime }\right) ^{2}+\left( \eta -\eta
^{\prime }\right) ^{2}+\left( \zeta -\zeta ^{\prime }\right) ^{2}}}\simeq
1.87.
\end{equation}
\bigskip
This value has been obtained by means of two different numerical methods
with the software @Mathematica.







\end{appendices}

\end{document}